\documentclass[10pt,journal,compsoc]{IEEEtran}
\ifCLASSOPTIONcompsoc
\usepackage[nocompress]{cite}
\else
\usepackage{cite}
\fi
\ifCLASSINFOpdf
\else
\fi
\usepackage[english]{babel}
\usepackage{amsmath}
\usepackage{amsfonts}
\usepackage{array}
\usepackage{graphicx}
\usepackage{bm}
\usepackage{fixmath}
\usepackage{blindtext}
\usepackage{amsthm}
\usepackage{mathrsfs}
\usepackage{enumerate}
\usepackage{amsthm}
\usepackage{float}\usepackage{nomencl}\usepackage{multirow}
\usepackage{bm}
\usepackage{lipsum}
\usepackage{amssymb}
\usepackage{algpseudocode,algorithm,algorithmicx}
\usepackage{epstopdf}
\usepackage{caption}
\usepackage{subcaption}\usepackage{xcolor}
\usepackage{url}\usepackage{lettrine}
\usepackage{balance}
\usepackage{enumerate}
\usepackage{amsmath}\usepackage{amssymb}
\usepackage{amsthm}
\usepackage{algpseudocode}
\usepackage{algorithm}
\usepackage{graphicx}
\usepackage{float}
\usepackage{bm} 
\usepackage{epstopdf}  
\usepackage{mathrsfs} \usepackage{upgreek}
\usepackage{mathrsfs}
\usepackage{enumerate}
\usepackage{amsmath}
\usepackage{amsthm}
\usepackage{algorithm,algpseudocode}
\usepackage{caption}
\usepackage{graphicx}\usepackage{lettrine}
\usepackage{float}
\usepackage{bm}\usepackage{array}
\usepackage{algorithm}
\usepackage{algpseudocode}
\usepackage{lipsum}
\usepackage{subcaption}
\usepackage{epstopdf} \usepackage{makecell}
\usepackage{nomencl}\usepackage{stfloats}
\usepackage{url}\usepackage{booktabs}\usepackage{colortbl}\usepackage{epstopdf}
\usepackage{etoolbox}
\def\BibTeX{{\rm B\kern-.05em{\sc i\kern-.025em b}\kern-.08em
		T\kern-.1667em\lower.7ex\hbox{E}\kern-.125emX}}

\newtheorem*{proof*}{Proof}

\newtheorem{remark}{Remark}

\definecolor{myblue}{RGB}{0,114,187}

\def\BibTeX{{\rm B\kern-.05em{\sc i\kern-.025em b}\kern-.08em
    T\kern-.1667em\lower.7ex\hbox{E}\kern-.125emX}}

\title{Imitation Learning for Adaptive Video Streaming with Future  Adversarial Information Bottleneck Principle}

\author{  
	\IEEEauthorblockN{Shuoyao Wang,~\IEEEmembership{Member, IEEE}}, \IEEEauthorblockN{Jiawei Lin}, and \IEEEauthorblockN{Fangwei Ye,~\IEEEmembership{Member, IEEE}}
	\thanks{Shuoyao Wang and Jiawei Lin are with the College of Electronic and Information Engineering, Shenzhen University, China (e-mail: sywang@szu.edu.cn; linjiawei2021@email.szu.edu.cn). Fangwei Ye is with  Broad Institute of MIT and Harvard, USA (e-mail: yefangwei@hotmail.com).}
	}

\begin{document}

\IEEEtitleabstractindextext{
\begin{abstract}
Adaptive video streaming plays a crucial role in ensuring high-quality video streaming services
. Despite extensive research efforts devoted to Adaptive BitRate (ABR) techniques, the current reinforcement learning (RL)-based ABR algorithms   may benefit the average Quality of Experience (QoE) but suffers from fluctuating performance in individual video sessions.
In this paper, we present a novel approach that combines imitation learning with the information bottleneck technique, to 
 learn from the complex offline optimal scenario rather than inefficient exploration.
In particular, we leverage the deterministic offline bitrate optimization problem with the future throughput realization as the expert and formulate it as a mixed-integer non-linear programming (MINLP) problem. To enable large-scale training for improved performance, we propose an alternative optimization algorithm that efficiently solves the MINLP problem. 
To address the issues of overfitting due to the future information leakage in MINLP, we incorporate an adversarial information bottleneck framework. By compressing the video streaming state into a latent space, we retain only action-relevant information. Additionally, we introduce a  future adversarial  term to mitigate the influence of  future information leakage, where Model Prediction Control (MPC) policy  without any future information is employed as the adverse expert. 
Experimental results demonstrate the effectiveness of our proposed approach in significantly enhancing the quality of adaptive video streaming,  providing a  7.30\% average QoE improvement and a 30.01\% average ranking reduction.
\end{abstract}

\begin{IEEEkeywords}
adaptive video streaming, imitation learning, information bottleneck, mixed-integer non-linear programming.
\end{IEEEkeywords}
}
\maketitle

\IEEEraisesectionheading{\section{Introduction}}
\IEEEPARstart{W}{ith}
 the development of Ultra-High-Definition (UHD) video, Virtual Reality (VR), and other video
streaming applications, video traffic is skyrocketing. 
Currently, video traffic is estimated
to account for 71 percent of all mobile data
traffic, and this share is forecast to increase
to 80 percent in 2028.
Supporting massive Quality-of-Experience (QoE) streaming has emerged as a crucial priority for video service providers in recent times.

Dynamic Adaptive Streaming over HTTP (DASH) stands out as a promising technology to efficiently deliver video content across diverse network conditions. Within DASH, the source video undergoes segmentation into multiple equal-length chunks, each of which is encoded at various bitrate levels. To ensure seamless playback, the client-side player employs an Adaptive BiRrate (ABR) algorithm, which dynamically selects the most suitable bitrate version for each chunk based on real-time estimations of the network conditions and current buffer occupancy.

Extensive research efforts have been devoted to ABR algorithm for improving user experiences. For example, 
heuristic-based ABR uses control rules to make bitrate decisions based on estimated
network \cite{10.1145/2934872.2934898}, 
and playback buffer size \cite{7524428},
or formulated optimization problem \cite{10.1145/2785956.2787486}. 
However, such methods introduce excessive  parameters that
need  carefully fine-tuning,  which result in 
unstable performance across  network conditions and QoE preferences.
By contrary, reinforcement learning (RL)-based ABR \cite{10.1145/3098822.3098843} \nocite{8526814}\nocite{8737418}\nocite{10.1145/3386290.3396930}-\cite{9334431} learns the strategy  in a data-driven manner, without using any 
pre-programmed control rules or explicit assumptions about the operating environment. 
Despite the flexibility and effectiveness of the DRL-based ABR algorithms, 
it is \emph{unclear whether} we have truly \emph{maximized} our efforts to deliver a \emph{high Quality of Experience (QoE)} ABR policy under limited downlink conditions.  
{Specifically, we  demonstrate in Section~3 that current ABR algorithms based on DRL may benefit the average Quality of Experience (QoE) but suffers from fluctuating performance in individual video sessions.}

Instead of  iteratively exploring and adapting actions to maximize rewards (i.e., average QoE), \cite{10.1145/3343031.3351014}\cite{9699071}   leverage
Imitation Learning (IL) to train the Neural Network (NN), where  the near-optimal policy, i.e., expert demonstration,  is instantly estimated via formulating the bitrate adaptation optimization problem with the future channel realizations. By imitating expert behavior, IL can avoid extensive exploration in reinforcement learning, and converge to complex but reliable policies quickly.
Unfortunately, the existing approaches still suffer the following challenges, hindering the advantages of IL.
\begin{itemize}
 \item \textbf{Heavy Overhead}: IL relies on high-quality expert demonstrations, which requires  significant computation or human feedback overhead  to obtain.  Existing efforts approximate the offline
 problem by truncated optimization with a limited future
 horizon, resulting in sub-optimal expert demonstration. 
 \item \textbf{Overfitting}:  Due to the fact that the  experts in ABR  utilizes the future realization in the training data,  
  the learned policies are more easily to be overfitted and thus  not generalize well during deployment.
\end{itemize}

In this paper, we propose ABABR (Adversarial Bottleneck ABR), a novel ABR algorithm based on imitation learning, which is scalable and can provide high QoE streaming service under unseen dynamics.  
In particular, we propose \emph{an alternative optimization algorithm to efficiently solve the multi-stage mixed-integer programming (MINLP) problem for  expert demonstration},  which is more than 3.7x faster than the conventional dynamic programming solver used in \cite{10.1145/3343031.3351014}\footnote{https://github.com/thu-media/Comyco}.
Moreover, to improve the generalization ability under imitation learning manner, we leverage \emph{information bottleneck (IB) principle, and propose a future adversarial  IB framework}.              
Our formulation aims at maximizing i) the mutual information between the expert action and the encoded feature, meanwhile, minimizing ii) the mutual information between the encoded feature and the current state, and iii) the future adversarial mutual information. Thus, it addresses the objectives of approaching the expert demonstration, while compressing the latent space representation and thus less overfitting, respectively.	Significantly, to the best of our knowledge, this is the first time that the IB technique has been applied to adaptive video streaming problems.  The main contributions of this work can be summarized as:
\begin{itemize}
	\item We propose ABABR, an innovative ABR algorithm combining imitation learning with the IB technique, capable of providing high-quality streaming service, in terms of \emph{both trace-average QoE and trace-wise ranking points}.

	\item  We  leverage the deterministic offline bitrate optimization problem with the
	future throughput realization as the expert, formulating it as an MINLP problem. To reduce the computation overhead, we propose an alternative optimization  algorithm that efficiently solves the MINLP problem, thus facilitating large-scale training for improved performance.
	
	\item To tackle the issues of overfitting  due to the future information leakage in MINLP, we introduce an advanced IB framework. By compressing the video streaming state into a latent space, the method retains only information relevant to action-making. A future adversarial term is further proposed to reduce the influence of future  information leakage,  where MPC policy without any future information is employed as the adverse expert.
	
	\item 
	{Trace-driven evaluation} results illustrate that ABABR can provide  an extra 7.30\% average
	QoE and a 30.01\% average ranking reduction.
\end{itemize}

The rest of the paper is organized as follows.
Section~2 and~3 summary the related work and the motivation of this work, respectively. 
We introduce the  problem formulation in Section~4. 
Section~5 elaborates the implementation of expert demonstration, adverse expert demonstration as well as the IB principle. 
In Section~6, experimental results and performance analysis
are presented. Finally, the paper is concluded in Section~7. 
For notation, we denote random variables by capital letters (e.g., $S$ and $A$) and their realizations by lower case letters (e.g., $s$ and $a$).  We denote by $H(S)$ and $I(S;A)$ as  the Shannon entropy of variable $S$ and the mutual information between $S$ and $A$, respectively.
\section{Related Work}
\subsection{Adaptive Video Streaming}
As one of the most frequently launched applications in wireless networks, ABR streaming service has attracted a great deal of attention. 
Existing approaches are broadly classified into three categories: heuristic-based strategies,  Model Predictive Control (MPC)-based strategies, 
and Reinforcement Learning (RL)-based strategies. 

\subsubsection{Heuristic}Several heuristic-based ABR schemes have been proposed by researchers, such as buffer-controlling \cite{7524428}, which regulates buffer occupancy to avoid re-buffering.
However, maintaining stable buffer occupancy may sacrifice bitrate utility, especially in dynamic networks. 
Combining rate-matching and buffer-controlling, BBA+ \cite{9762785} proposes a non-linear throughput-aware mapping function that projects the buffer occupancy observation to the bitrate decision. 
To make the system more generalizable to different types of unforeseen environments, Cratus \cite{9312489} proposes an light-weighted online algorithm to control the buffer dynamic behavior. 
Unfortunately, these schemes cannot balance various QoE factors, such as bitrate utility and smoothness.

\subsubsection{Model Predictive Control}To balance various QoE factors, numerous research efforts have focused on MPC algorithms. In particular, MPC-based ABR algorithms predict throughput when downloading future chunks using a sliding window.
For instance, RobustMPC \cite{10.1145/2785956.2787486}
employs the harmonic mean of previous throughput samples as the throughput prediction, and maximize the long-term QoE by taking the prediction as the exact future realization. 
Afterwards, many research efforts have been devoted to improving the time-slotted throughput prediction for MPC, such as  CS2P \cite{10.1145/2934872.2934898} and  Oboe \cite{10.1145/3230543.3230558}. 
Moreover, \cite{9796948}-\cite{10041952} further improved throughput prediction across various networks by integrating physical and link layer information. 

Fugu \cite{yan2020learning} and PAR \cite{9860000} propose to predict chunk-average throughput, rather than time-slotted throughput, to more accurately predict future buffer occupancy. 
Considering the difficult to predict future network throughput accurately, \cite{kan2021uncertainty} leverages Bayesian Neural Network to predict the probability distribution of future throughput, rather than the throughput.
However, wireless channel conditions can change rapidly in 5G, 
making throughput prediction inaccurate, even the distribution. 
Furthermore, the above methods  introduce additional hyper-parameters, 
leading to inconsistent performance in different networks. 

\subsubsection{Reinforcement Learning}Inspired by the sky-rocketing development of the deep learning, learning-based algorithm has recently appeared as a promising solution to ABR streaming problem,  
without using
any pre-programmed control rules or explicit assumptions
about the operating environment. 
For example, Pensieve \cite{10.1145/3098822.3098843}  proposed  Asynchronous Advantage Actor Critic (A3C)-based ABR schemes to maximize the users' QoE on video streaming.
Afterwards, extensive research efforts have been devoted to employing advanced machine learning algorithms to better make the bitrate selection decision, such as cascaded RL \cite{8526814},   Bayesian bandits \cite{8737418}, imitation learning \cite{10.1145/3343031.3351014}, self-play RL \cite{10.1145/3386290.3396930}, Proximal Policy Optimization \cite{ICME23}, and Phasic Policy Gradient \cite{tianchihuang}.
Considering the deployment cost and environment adaptation problem, \cite{9334431} proposes Pitree to distill the ABR policy and \cite{9796516}\nocite{10.1145/3503161.3548331}-\cite{10077780}  adopted  meta-RL-based ABR systems  to improve generalization ability among various networks, respectively. 
Nevertheless, most of the existing RL method require massive interactions with the environment, which is time-consuming and computational expensive. Moreover, it is unclear whether we have truly maximized our efforts to
deliver a high QoE ABR policy under
limited downlink conditions   
\subsection{Imitation Learning}
Imitation learning (IL) is proposed to enable intelligent agents by imitating expert behaviors for  decision-making. 
In the IL  setting, we are given a set of expert
trajectories, with the goal of learning a policy which induces behavior similar to the expert’s. 

The most traditional approach to IL is Behavioral Cloning (BC) \cite{ross2010efficient}, where a classifier or regressor is trained to fit the behaviors of the expert.
While BC works well in simple environments with massive data, 
it fails to learn a good policy in complex environments and suffers significant performance degradation on unseen states during deployment.
To improve the generalization performance of IL, many research efforts have been devoted, such as data augmentation \cite{ross2011reduction},  adversarial IL \cite{Yang_Vereshchaka_Zhou_Chen_Dong_2020}, and learning rewards with imitation \cite{garg2021iq}.
However, 
most of the above mentioned work ignore the
sequential nature of the decision-making problem, and small errors can accumulate and quickly compound when the
learned policy departs from the states observed under the expert.
Alternatively, apprenticeship IL is proposed to allowing the learner to further interactive with the environment, a.k.a, imitative deep reinforcement learning \cite{Liu_Liu_Zhao_Pan_Liu_2020}\cite{singh2021parrot}. Unfortunately, to enable an automatic interactive system, heuristic policies are most adopted to generate the expertise demonstration. The sub-optimal heuristic may leads to degraded learned policy.

The success of IL has sparked to the communication and networking community. Thanks to the powerful modeling method, the optimization algorithm, rather than the heuristic algorithm,  is in perfect position to generate the apprentice expertise demonstration. For instance, \cite{10032264} and \cite{10296872} leverage BC to optimize   360-degree video transmission and live video analytics with optimization-based experts, respectively. 
To enhance the generality, \cite{10026241} proposed a match-and-update offline optimization algorithm as  the expert and leverage DAgger \cite{ross2011reduction} to learn an online agent to schedule services and  manage resources, where the strategy obtained by cloning the behavior further interacts with the environment to generate new data and the new data is then labeled by the expert for the next round BC. 
\subsubsection{Imitation Learning-based ABR}
In terms of adaptive video streaming, the dense research efforts on model-based approach like MPC \cite{10.1145/3230543.3230558}-\cite{kan2021uncertainty}  can give remarkable bitrate selection under the
current state with the clear future network information during the offline training phase. Therefore, MERINA~\cite{10.1145/3503161.3548331} adopt RobustMPC~\cite{10.1145/2785956.2787486} as the expert to pretrain the actor network in an IL manner, to cope with the slow convergence problem in meta RL. Utilizing the future throughput realization,   \cite{10.1145/3343031.3351014} \cite{9699071} adopt truncated offline optimization with dynamic programming solver as the apprentice expert to expedite the learning process of ABR agent. In most recent, \cite{LianchenJia} show that the algorithms can also utilized for joint optimization of resolutions and duration
to get resolutions-duration ladders that maximize the video streaming QoE. 

However, the above approaches still suffers high computation time and easy overfitting problem. Specifically, 
 \cite{10.1145/3343031.3351014} \cite{9699071}    approximate the offline problem by truncated
 optimization with a limited future horizon, resulting
 in sub-optimal expert demonstration.
Moreover, the offline optimization-based experts also integrate the further throughput information in the training data, thereby leading the demonstration and thus the learned agent overfit to the training data. The agents suffer significant performance degradation  on unseen states during
testing and deployment.  The above challenges may hinder the advantages of IL in existing studies. 

\section{Background and Motivation}
In this section, we provide a concise overview of the specific
scenario under consideration. We then use empirical measurements to elucidate the key limitations of prior solutions.
 \begin{figure}[!t]
	\centering
	\includegraphics[clip, trim={0cm 0.1cm 0cm 0cm}, width=0.95 \linewidth]{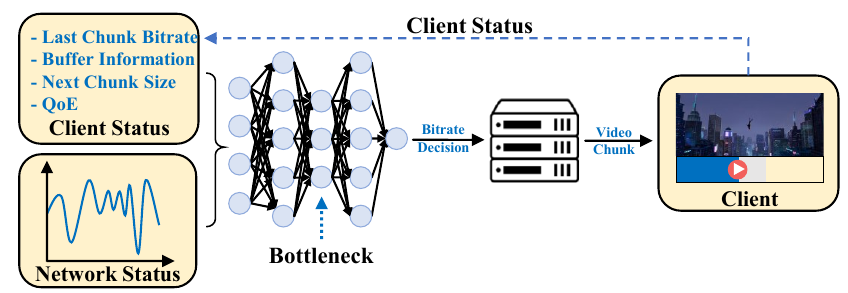}
	\caption{\small Information Bottleneck-enabled streaming system.}
	\label{sys}
\end{figure}
\subsection{Adaptive Video Streaming}
As shown in Fig.~\ref{sys}, we consider a DASH system where a source video is split into $I$ chunks, and each of which contains $L$ seconds of video content. 
Each chunk is encoded into multiple copies with different bitrates. The ABR algorithm  select the appropriate version of the chunk, to ensure a high user-of-experience streaming service.
For notation simplicity, we denote the finite set of bitrate levels
as $\mathcal{R}  = \{R_1, R_2, ..., R_{|\mathcal{R}|}\}$.
Without loss of generality, 
we assume that $R_1 > R_2 > ... > R_{|\mathcal{R}|}$,
and denote $r_i \in \mathcal{R}$ as the bitrate level of chunk $i=1,\dots,I$.

In particular, the client downloads the video chunks via a wireless downlink, 
and the downloading time $\tau_i$ of chunk $i$ can be calculated as:
\begin{equation}
	\tau_i = \frac{s_i(r)}{\frac{1}{\tau_i} \int_{t_i}^{t_i+\tau_i} c_{t} \,dt },  
	\label{time}
\end{equation} 
where $s_i(r)$ is the data volume of the copy with bitrate $r$, 
$t_i$ is the start time of downloading the $i$-th chunk, and
$c_{t}$ is the downlink bandwidth at time $t$.

To enable a smooth playback, a buffer is drained as the user watches the video 
and is filled with $L$ seconds whenever a chunk is downloaded. The buffer occupancy $b_i$ describes buffered video time when player starts
downloading chunk $i$: 
\begin{equation}
	b_{i+1}={[b_i - \tau_i]}^+ + L, 
	\label{buffer}
\end{equation}
where $b_i \in [0, B]$, and
$B$ denotes the buffer capacity. 
If $b_{i+1} - \tau_i<0$, the buffer is exhausted  and 
the player has to keep re-buffering until a new chunk is fully downloaded.

The ABR algorithm aims to generate a policy $\pi$ to maximize the QoE of the entire
video session. There are many different definitions of QoE, such as quality-aware QoE 
metric \cite{10.1145/3343031.3351014}, ITU-T Rec. P.1203. 
For direct comparison with fundamental ABR algorithm, we employ the
cumulative QoE objective function used by \cite{10.1145/3098822.3098843}:

\begin{equation}
	\text{QoE} = \sum_{i=1}^{I}q(r_i)
	- \alpha_1 \sum_{i=1}^{I}[\tau_i - b_i]^+
	- \alpha_2 \sum_{i=2}^{I} \left\lvert {q(r_{i}) - q(r_{i-1})} \right\rvert,
	\label{qoe}
\end{equation}
where
$q(r_i)$ represents the video quality function of the $r_i$ bitrate copy of chunk $i$.
The first term indicates that higher bitrates help improve QoE.
The second term represents re-buffer  impairs the QoE. 
The variation penalty is calculated in the third term.     
The symbols $\alpha_1$ and $\alpha_2$ are preference weights for 
re-buffer and variation penalties, respectively.

\begin{figure}[!t]
	\centering
	\includegraphics[clip, trim={1.5cm 0.15cm 2cm 0.5cm}, width=1.0\linewidth]{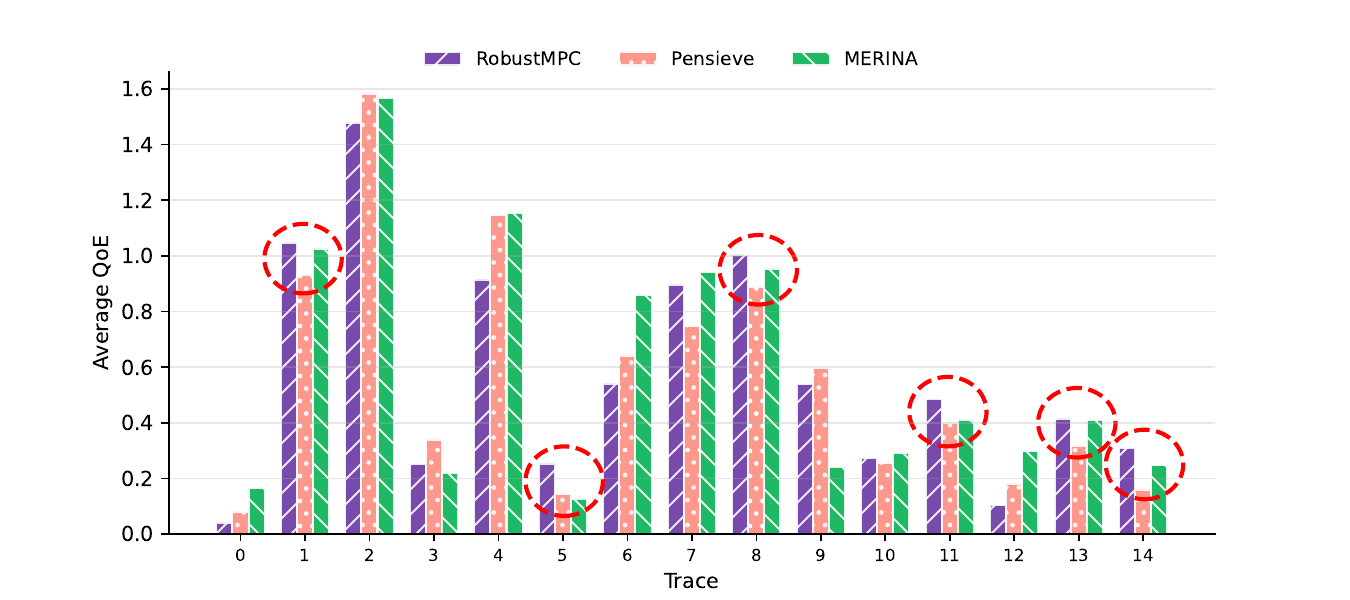}
	\caption{\small The average QoE of different algorithms under different traces. The red circles indicate the video session that RobustMPC outperforms the RL methods. }
	\label{fig:trace-level}
\end{figure}
\subsection{Challenges and Motivations}
In this section, we answer the following question: 
\textit{Why do we study imitation learning based ABR algorithms?}
Many research efforts have been devoted to utilizing reinforcement learning (RL), 
including meta-RL, to enhance ABR system performance and stability. 
{
These endeavors have resulted in significant enhancements in average QoE, 
exemplified by Pensieve~\cite{10.1145/3098822.3098843} and MERINA~\cite{10.1145/3503161.3548331}, 
showcasing an improvement of 5.4\% and 8.6\% over RobustMPC\cite{10.1145/2785956.2787486}.
However, it is still questionable that whether we have done our best to provide high QoE ABR policy 
under limited downlink conditions. 
We evaluate the representative ABR algorithms across diverse network traces follow as 
\cite{10.1145/2785956.2787486}\cite{10.1145/3098822.3098843}\cite{10.1145/3503161.3548331}, and the 
average QoE is computed over the entire video session of each trace. 
As shown in Fig~\ref{fig:trace-level}, we find that both Pensieve and MERINA suffer from fluctuation performance in individual video sessions. In particular, we observe that 
According to \cite{10258330} and \cite{pmlr-v119-siddique20a}, 
RL works by maximizing cumulative rewards over time through interaction with the environment. 
This process results in the development of a high-reward policy across the entire training dataset. 
However, the adoption of such a policy may come at the cost of compromising 
the individual performance of specific traces.}

\begin{figure}[!t]
	\centering
	\includegraphics[clip, trim={1.2cm 0.2cm 3cm 0.2cm}, width=0.85\linewidth]{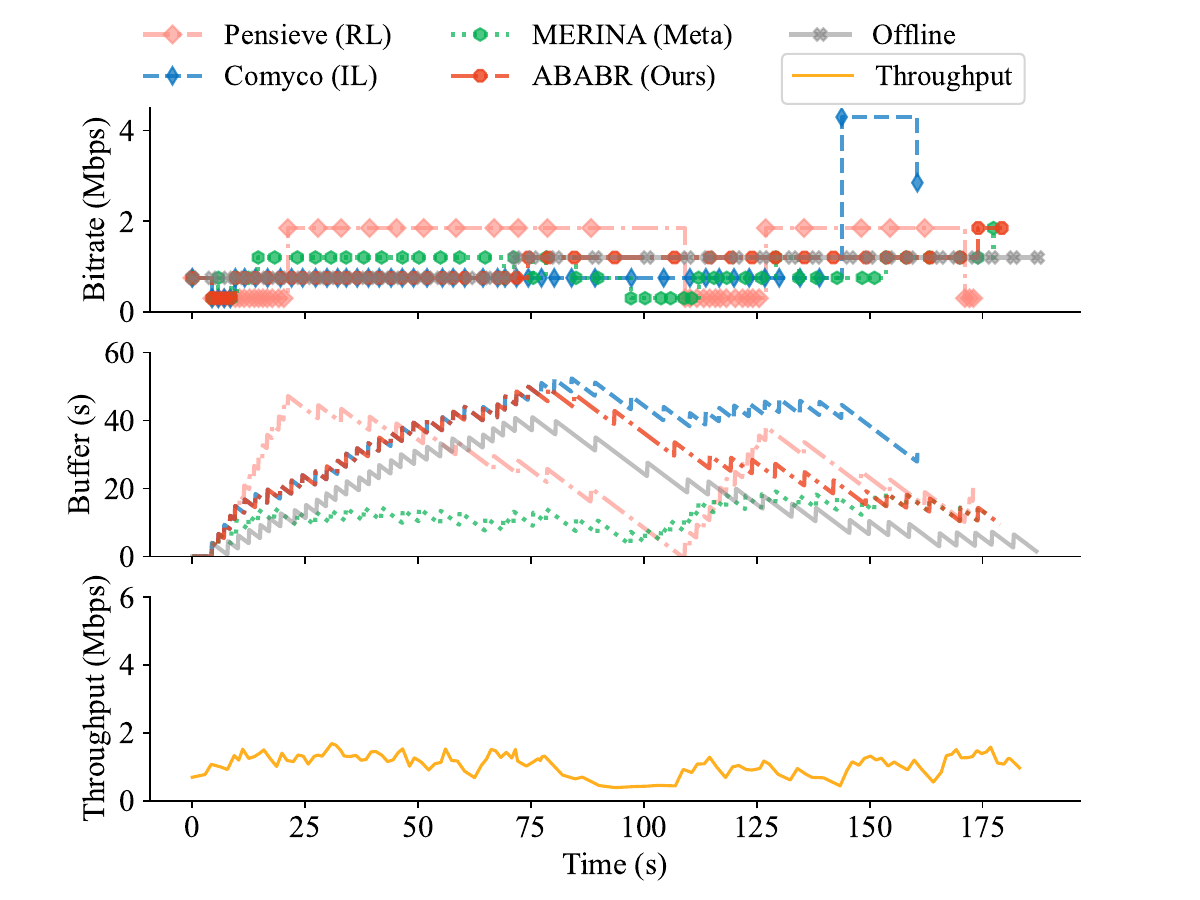}
	\caption{\small An example of various algorithmic policies.}
	\label{fig:motivation}
\end{figure}

Moreover, 
we plot the bitrate selections implemented by the existing policies as well as the offline optimal solutions 
in Fig.~\ref{fig:motivation}, to identify gaps between the existing ABR policies and optimal solutions.
As shown in Fig.~\ref{fig:motivation}, the offline mixed-integer nonlinear programming (MINLP) solution exhibits a preference   for selecting video chunks with smaller bitrate during the early stages of the video session,prioritize building a sufficient buffer size.
Notably, an adequate buffer size can prevent rebuffering or rapid bitrate degradation when the network throughput falls down.   As the buffer size reaches a threshold, adjustments are made based on
considerations of buffer size and prevailing network conditions.
In contrast, the
RL-based approach MERINA and Pensieve pays little attention to
buffer size accumulation, making it more \emph{susceptible to
network fluctuations}.

In addition,  \emph{having a higher average QoE across various video sessions does not necessarily translate to an overall higher satisfaction} \cite{10.1145/3386290.3396930}. Accordingly, we evaluate trace-wise ranking QoE comparisons in Section~6.1.2. While RL-based methods achieve a remarkable average QoE by preventing bad cases through extensive exploration, they seldom attain the best QoE in each trace. Given the fact that the QoE model is a relative score, we consider trace-wise ranking performance as a crucial metric, where the proposed ABABR demonstrates more than a 23.09\% ranking point improvement and 30.01\% average ranking reduction.  
Indeed, IL-based methods, such as Comyco and ABABR in Fig.~\ref{fig:motivation}, can learn 
the policies according to the expert demonstrations instead of exploring in the environment, 
making it more stable than others. 
In summary, IL-based methods excel in training efficiency, stability, average QoE performance, and, most importantly, trace-wise QoE ranking performance.

Hence, in this paper, we concentrate on addressing the heavy overhead and unseen degradation issues associated with existing IL approaches. We propose an alternative optimization method to efficiently solve the MINLP problem in Section 5.3, addressing the heavy overhead. Additionally, we employ the information bottleneck technique and introduce the adverse expert in Section 5.4. Together with variational approximation in Section 5.5, these methods effectively prevent the overfitting problem of IL-based ABR, thereby enhancing the user experience.

\section{Problem Formulation}


\subsection{Offline Problem Formulation}
In summary, the QoE maximization problem can be described as (P1):
\begin{subequations}
	\begin{align}
		\underset{\bm{r},\bm{\tau},\bm{t}}{\max} \quad &  \text{QoE},    \\
		\text{s.t.} \quad & r_i \in \mathcal{R}, \forall i,      \\
		& \tau_i = \frac{s_i(r)}{\frac{1}{\tau_i} \int_{t_i}^{t_i+\tau_i} c_{t} \,dt },\forall i,   \\
		&    b_{i+1}={[b_i - \tau_i]}^{+} + L,   \forall i,\\
		& t_{i+1}=t_{i}+\tau_i, \forall i, 
	\end{align}
\end{subequations}
where $\bm{r}=(r_1,\dots,r_I)$, $\bm{\tau}=(\tau_1,\dots,\tau_I)$, and $\bm{t}=(t_1,\dots,t_I)$. Notice that the objective of the ABR algorithm is to determine the sequential bitrate decision $\bm{r}$. However, the QoE in (\ref{qoe}) is also related to the downloading time. To enabling optimization, we introduce the intermediate decision variables $\bm{\tau}$ and $\bm{t}$ in (P1). This  formulates the downloading starting time $t_i$ and duration $\tau_i$ mathematically, preventing iteratively simulate the downloading time with  simulator.  Since the constraints are in equality  format, given $\bm{r}$, the solution of intermediate variables $\bm{\tau}$ and $\bm{t}$ are unique. 

Given the future information of $c_t$, (P1) is an MINLP  with nonlinear and undifferentiated constraints (4c-d),  which is typically time-consuming to obtain a (near)-optimal solution.



\subsection{MDP Formulation}
Unfortunately, the future downlink bandwidth $c_{t}$ is hard to predict, specifically in wireless networks.
In practice, based on information from the current and past environments (e.g., last chunk bitrate, re-buffer time), 
the ABR algorithm may only determines the next chunk bitrate sequentially to 
maximize the QoE of the entire video session. 
The sequentially decision-making problem can be naturally formulated as
a Markov decision process (MDP) problem. It involves five elements,
including environment state, action, transition function, reward function, and discount factor.

\emph{1) Action: } 
At the beginning of the downloading process of chunk $i$, the video streaming application determines the discrete bitrate adaptation decision. 
The action is defined as:
\begin{equation}
	\bm{a}_i = r_i \in \mathcal{R}.
	\label{action}   
\end{equation}

\emph{2) State: } 
The streaming system collects informative features as a
state, providing evidence for the ABR controller/agent   to take action.
In this section, we explicitly define informative features
in terms of fundamental properties of the problem.

First of all, due to the fact that the channel throughput has a strong temporal correlation, although hard to predict, the download time for the past $k$ chunks $\bm{\tau}^{i}=(\tau_{i-k+1},\dots,\tau_i)$ and  the network throughput measurement for the past $k$  chunks $\bm{p}^{i}=(\frac{n^{i-k+1}_{r_{i-k+1}}}{\tau_{i-k+1}},\dots,\frac{n^{i}_{r_{i}}}{\tau_{i}})$ are part of the state. Since the downloading time is also related to the downloading file size, the available chunk sizes for the next 
chunk $ \bm{n}^{i}=(n^{i}_{R_1},\dots,n^{i}_{R_{|\mathcal{R}|}})$ are included in the state.

As shown in QoE function (\ref{qoe}), the playback buffer occupancy $b_i$ and the last bitrate $r_{i-1}$ directly determine the re-buffering penalty and the fluctuation penalty, respectively. Therefore, $r_i$ and $b_i$
are also part of the current state. Moreover, to balance the current and future QoE, the number of remaining chunks $m_i$ is also important for QoE , and thus included in the state.          Overall, the agent determines the bitrate decision for each chunk $i$ 
according to the following observations, i.e., state $s_i$:
\begin{equation}
	\bm{s}_i = (\bm{p}^{i}, \bm{\tau}^{i}, \bm{n}^{i}, r_i, b_i, m_i).
	\label{state}
\end{equation}

\emph{3) Reward: }
The reward function is designed according to the objective
of the video streaming network. 
According to (\ref{qoe}), the streaming objective is the summation of QoE over each chunk. Therefore, we formulate the reward function at chunk $i$ as follows:
\begin{equation}
	v_i(\bm{s}_i, \bm{a}_i) = q(r_i)
	- \alpha_1{[\tau_i - b_i]}^+
	- \alpha_2 \left\lvert {q(r_{i}) - q(r_{i-1})} \right\rvert.
	\label{reward}
\end{equation}

Ultimately, the ABR agent aims to find the optimal policy 
$\pi$ by solving the following MDP problem:
\begin{equation}
	\begin{aligned}
	    \quad \max_{\pi} \qquad & V_{\pi}(\bm{s}_0) = \mathbb{E}
		\left[ \sum_{i=1}^I \gamma^i v_i(\bm{s}_i, \bm{a}_i)|\pi, \bm{s}_0 \right],     \\
		s.t. \qquad & \bm{a} \in \mathcal{A}, \forall \bm{a} \thicksim \pi(\bm{s}_i), \forall i ,
	\end{aligned}\label{mdp}
\end{equation}

\noindent where $\bm{s}_0$, $\pi: \mathcal{S}\rightarrow \mathcal{A}$, and $\gamma \in (0, 1]$
are the initial network state, the policy to map the state to actions, and the 
discount factor to balance instantaneous and future, respectively. 

\begin{remark}
	Many research efforts have been also devoted to i) using \emph{feature engineering on the application layer logs} in Eq.~(\ref{state}) (e.g., group layer normalization), ii) \emph{additional cross layer feature}, and iii) \emph{reward engineering},    to improve the performance and convergence speed of the learning algorithms. For fair comparison, we keep the state as simple as the state in \cite{10.1145/3098822.3098843} and thus be orthogonal to the feature/reward engineering as well as the meta-learning efforts. We can further improve our work from the efforts, which, however, is out of the scope of this paper.
\end{remark}

\section{Methodology}
\subsection{Actor Network}
Since the channel distribution in (\ref{mdp}) is  not easy to estimate precisely in a cost-effective manner,   we employ an actor network to parameterize the solution $\pi$ in (\ref{mdp}) with  $\bm{\bm{\theta}}$. Accordingly, we denote the parameterized policy as $\pi_{\bm{\theta}}(\cdot|s)$.
In conventional supervised learning manner, the parameter $\bm{\theta}$ is trained to minimize distance measurements, such as cross entropy and mean square error, between the network output and the ground truth label. However, in ABR system, the ground truth for the bitrate selection is generally missed, and thus policy gradient technique is widely used to train the actor network. 
 The RL agent policy is learned by minimizing the expectation of a intermediate value function: 
\begin{equation}
	L_R(\bm{\theta})=\mathbb{E}_{\bm{s},\bm{a}\sim\pi_{\bm{\theta}}(\bm{s})}[Q^{\pi}(\bm{s},\bm{a})],
\end{equation}
with
\begin{equation}\label{qfunc}
	Q^{\pi}(\bm{s},\bm{a})=\mathbb{E}^{\pi}\{ \sum_{i=1}^{m|\bm{s}} \gamma^i v_i(\bm{s}_i, \bm{a}_i)|\bm{s}_1=\bm{s},\bm{a}_1=\bm{a}\},
\end{equation}
where $m|\bm{s}$ denote the remaining chunks of this video session given the state $\bm{s}$. 
However, obtaining the intermediate Q-function in (\ref{qfunc}) itself is challenge. To cope with the unknown Q-function,  the actor-critic framework is proposed, where the learning process alternates
between policy improvement, i.e., actor module with parameter $\bm{\theta}$, and policy evaluation, i.e., critic module with parameter $\bm{\phi}$, in the direction of maximizing $Q^{\pi_{\bm{\theta}}}_{\phi}(\bm{s},\bm{a})$. Accordingly, the  ABR policy is learned by minimizing
\begin{equation}
	L_R(\bm{\theta})=\mathbb{E}_{s,a\sim\pi_{\bm{\theta}}(\bm{s})}[Q_{\bm{\phi}}(\bm{s},\bm{a})].
\end{equation}

\begin{figure}[!t]
	\centering
	\includegraphics[width=0.88\linewidth]{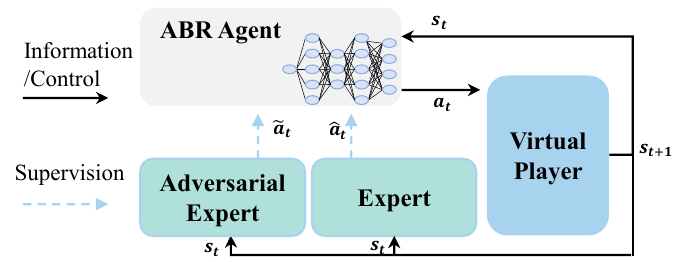}
	\caption{\small An overview of imitation learning framework. In this paper, Expert is achieved by the offline optimal with the alternative optimization algorithm in Section 5.3. Adverse Expert is achieved by the RobustMPC benchmark~\cite{10.1145/2785956.2787486}.}
	\label{il}
\end{figure}
Unfortunately, actor-critic algorithms, due to the adversarial training nature,  suffer from issues such as instability, sample inefficiency, sensitivity to hyper-parameters, and the potential for suboptimal convergence. 
In light of this, we have decided to employ imitation learning as an alternative approach to solving the adaptive video streaming problem. By using imitation learning, we aim to overcome the limitations of DRL and achieve more reliable and predictable outcomes. As shown in Fig.~\ref{il}, we leverage the expert demonstration as the label information and the imitation ABR policy is trained by minimizing
\begin{equation}
	L_I(\bm{\theta})=\mathbb{E}_{\bm{s},\bm{a}\sim\pi_{\bm{\theta}}(\bm{s})}\Big[D_{KL}\big(\pi_{\bm{\theta}}(\cdot|s)||\pi^{*}(\cdot|\bm{s})\big)\Big],
\end{equation}
where $\pi^{*}(\cdot|s)$ denotes the expert demonstration policy.  

\subsection{Expert Demonstration}
In this subsection, we discuss how to generate high-quality expert demonstration to 
obtain a high-perform imitated agent. The cost of hiring human experts to mark the optimal 
bit rate selection for ABR is huge. And due to the limited expert demonstration, the performance of the behavior cloning ABR policy is also limited.  Alternatively, as the model-based approach  give near-optimal bitrate selection under the
current state with the clear future network information during
the offline training phase, it is an excellent selection to apply
offline deterministic optimization as the apprentice expert to expedite the learning process of neural
network.

In particular, we formulate a QoE maximization problem for $N$ consecutive chunks, 
i.e., the current and the next $N-1$ chunks, with future network throughput measurement $c_t$. The future realization can be
successfully collected under both offline environments and
real-world network scenarios. Inspired by (P1), 
the optimization problem of the $N$ consecutive chunks when making the decision for 
chunk $i$ can be described as (P2):
\begin{equation}
	\begin{aligned}
		\underset{\bm{r},\bm{\tau},\bm{t}}{\max}  \;& \sum_{\iota=i}^{i+N-1} \big[q(r_\iota)
		- \alpha_1  {[\tau_\iota -b_\iota]}^+ 
		- \alpha_2  \left\lvert {q(r_{\iota}) - q(r_{\iota-1})} \right\rvert\big],    \\
		\text{s.t.}  \; & r_\iota \in \mathcal{R}, \;\; \iota=i,\dots,i+N-1,      \\
		& \tau_\iota = \frac{s_\iota(r)}{\frac{1}{\tau_\iota} \int_{t_\iota}^{t_\iota+\tau_\iota} c_{t} \,dt },\;\; \iota=i,\dots,i+N-1,   \\
		&    b_{\iota+1}={[b_\iota - \tau_\iota]}^{+} + L,  \;\;  \iota=i,\dots,i+N-1, \\
		& t_{\iota+1}=t_{\iota}+\tau_\iota,\;\; \iota=i,\dots,i+N-1. 
	\end{aligned}
\end{equation}
Unfortunately,  (P2) is still a non-convex MINLP, and thus  dynamic programming method introduces significant computation overhead, especially for large $N$. On the other hand, existing works have shown that a sufficiently large $N$ is necessary to improve the expert demonstration. Ideally, there
exists a trade-off between the computation overhead and the expert performance.

	\subsection{Alternative Optimization}

	 To reduce the computation overhead of solving the MINLP, in the following, we decompose (P2) into Chunk-Average Throughput-enabled QoE Maximization Problem and Chunk-Average Throughput Estimation Problem. 
	Then, we solve the two problems iteratively. This process repeats until  
	the throughput estimation donnot changes.  Since the proposed alternative optimization is not a recursive loop that requires complex dynamic programming, and the branch-and-bound algorithm has a wealth of open source acceleration solvers, the proposed method greatly reduces the computation time compared with dynamic programming, as shown in Section~6.2. The detailed decomposition are as follows.

For notation simplicity, we set the chunk index $i=1$ in this section.  We denote by $\bar{c}_j$ as the  average throughput  when downloading chunk $j$. Then, (P2) can also be expressed as:
\begin{subequations}\label{P3}
	\begin{align}
		\underset{\bm{r},\bm{\tau},\bm{t}}{\max}  & \sum_{j=1,\dots,N} \Big[q(r_j)
		- \alpha_1  {[\tau_j - b_j]}^+
		- \alpha_2  \left\lvert {q(r_{j}) - q(r_{j-1})} \right\rvert\Big],    \\
		\text{s.t.}\; & r_j \in \mathcal{R}, \forall j=1,\dots,N,      \\
		& \tau_j = \frac{s_j(r)}{\bar{c}_j},\forall j=1,\dots,N,   \\
		&    t_{j+1}=t_{j}+\tau_j,\forall j=1,\dots,N,\\
		&    b_{j+1}={[b_j - \tau_j]}^{+} + L,   \forall j=1,\dots,N, \\
		& \bar{c}_j =\frac{1}{\tau_j} \int_{t_j}^{t_j+\tau_j} c_{t} \,dt,\forall j=1,\dots,N. 
	\end{align}
\end{subequations}
\subsubsection{QoE Maximization Problem Given Throughput}
We note that the bitrate selection problem given $\bar{c}_1,\dots,\bar{c}_N$ is a simpler mixed-integer programming problem with linear constraints. In particular, given the estimated $\bar{c}_1,\dots,\bar{c}_N$, we can express the bitrate selection problem as:
\begin{equation}
	\begin{aligned}
		\underset{\bm{r},\bm{\tau},\bm{t}}{\max}  & \sum_{j=1,\dots,N} \Big[q(r_j)
		- \alpha_1  {[\tau_j - b_j]}^+
		- \alpha_2  \left\lvert {q(r_{j}) - q(r_{j-1})} \right\rvert\Big],    \\
		\text{s.t.} \;& \text{Constraints}\;(\ref{P3}\text{b-e}).
	\end{aligned}\label{MINLP}
\end{equation}
To cope with the non-linearity of ${[\tau_j - b_j]}^+$, we denote by $e_i$ as the re-buffering time at chunk $i$. Accordingly, solving (\ref{MINLP}) is equivalent to finding the optimal decisions for the following problem:
\begin{subequations}\label{P5}
	\begin{align}
		\underset{\bm{r},\bm{\tau},\bm{t},\bm{e}}{\max}  & \sum_{j=1,\dots,N} \Big[q(r_j)
		- \alpha_1  e_j
		- \alpha_2  \left\lvert {q(r_{j}) - q(r_{j-1})} \right\rvert\Big],    \\
		\text{s.t.} \; & \text{Constraints}\;(\ref{P3}\text{b-d}),      \\
		&    b_{j+1}=b_j - \tau_j+ e_j+ L,   \forall j=1,\dots,N,\\
		& e_j \geq 0, \forall j=1,\dots,N,\\
		& e_j \geq \tau_j - b_j, \forall j=1,\dots,N.
	\end{align}\label{MINLP2}
\end{subequations}
\noindent where the objective (\ref{MINLP2}a) and the constraints (\ref{MINLP2}c-g) are linear. Therefore,  Problem (\ref{MINLP2}) is a mixed-integer linear programming (MILP), and thus can be efficiently calculated with a standard MILP algorithm, e.g., branch-and-bound algorithm. 
\subsubsection{Chunk-Average Throughput Estimation Problem}
As discussed in \cite{9860000}, the bitrate decision itself also influences the chunk-average throughput, since the different downloading file size indicates different downloading period each at the same throughput trace. Therefore, after we obtain the new optimal  bitrate from (\ref{MINLP2}), the new bitrate may mismatches with the last estimated throughput $\bar{c}_1,\dots,\bar{c}_N$. Therefore, we need to re-estimate the chunk-average throughput. In particular, given the current bitrate selection $\bm{r}^*$, we formulate the chunk-average throughput estimation problem as the re-buffering minimization problem:  
\begin{subequations}
	\begin{align}
		\underset{\bm{\tau},\bm{t},\bm{e},\bar{\bm{c}}}{\max} \qquad & \sum_{j=1}^{N} 
		- \alpha_1  e_j,    \\
		\text{s.t.} \qquad 
		& \text{Constraints}\;(\ref{P5}\text{b-e}),  \\
		&\bar{c}_j =\frac{1}{\tau_j} \int_{t_j}^{t_j+\tau_j} c_{t} \,dt,\forall j=1,\dots,N.   
	\end{align}\label{CT}
\end{subequations}After solving (\ref{CT}), we obtain the chunk-average throughput under the optimal downloading start time $\bm{t}^{*}$ and  duration $\bm{\tau}^{*}$, denoted as $\bar{c}^{*}$. If the updated throughput is equal to the last estimation, i.e., $\bar{c}^{*}==\bar{c}$, we conclude that the current solution  $(\bm{r}^{*},\bm{t}^{*}, \bm{\tau}^{*})$ is a feasible solution to (P2), and thus terminate the algorithm. Otherwise, we update the estimation with $\bar{c} \gets \bar{c}^{*}$ and return a new (\ref{P3}).

\subsection{Information-Bottleneck in Actor Network}
\begin{figure}[!t]
	\centering
	\includegraphics[clip,  trim={0cm 0cm 0cm 0cm},width=0.85 \linewidth]{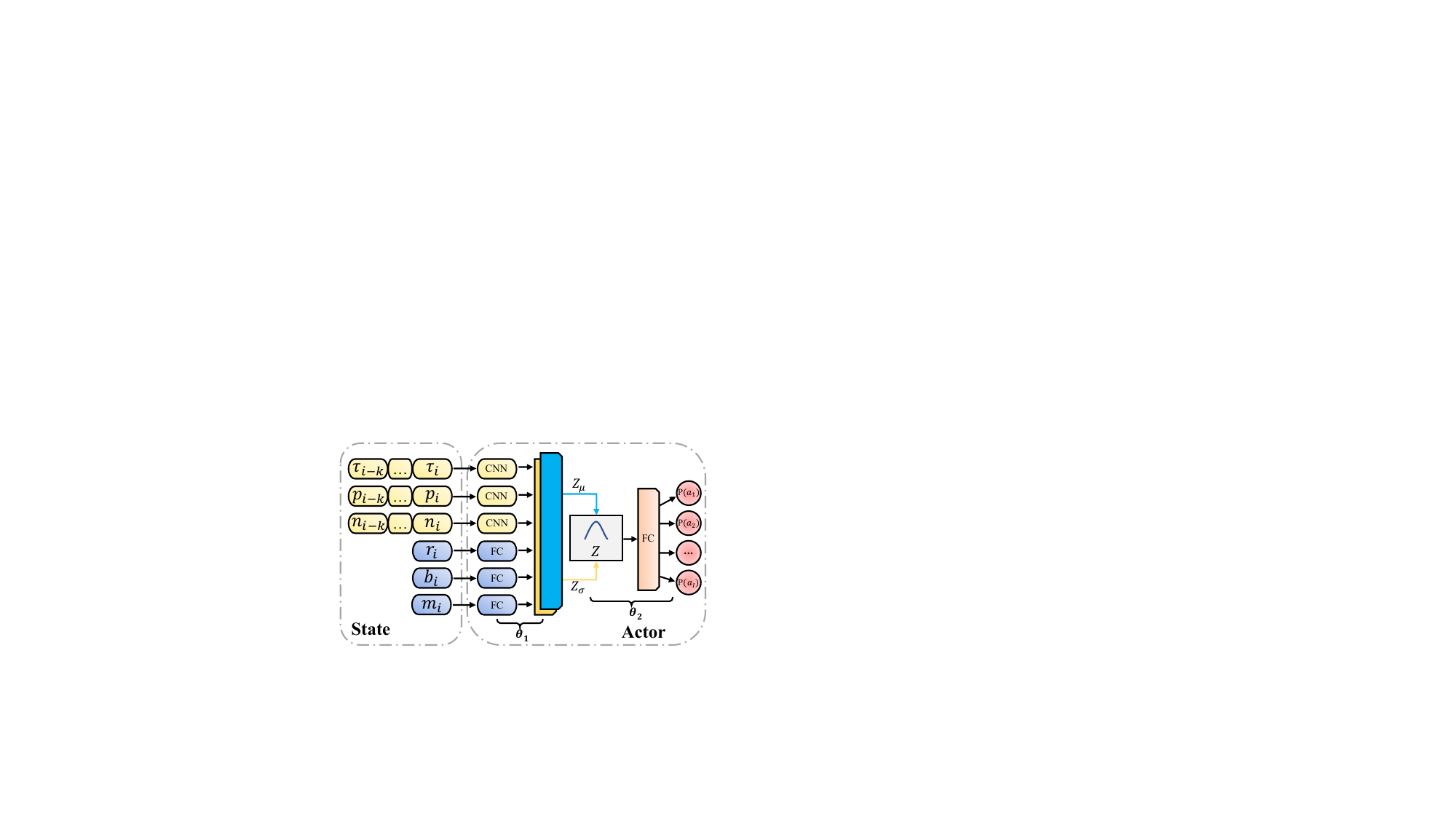}
	\caption{\small Actor Network with Information-Bottleneck.}
	\label{actor}
\end{figure}

{For the actor network, we first use the CNN layers and fully connected layers to process different types of data in state.} 
In the conventional imitation learning-based ABR algorithms \cite{10.1145/3343031.3351014}\cite{9699071}\cite{LianchenJia}, the actor network improves the learned policy by directly minimize the distance between the expert demonstration and the  learned policy, a.k.a, behavior cloning. However, the learned policy can be overfitting and thus degrades significantly for unseen scenarios, since the offline optimization algorithm leaks the future channel information in the training data and directly behavior cloning maybe lead the agent to memorize future network situations rather than making reasonable inference based on the current state. To tackle this issue, we decompose the neural network parameterized with $\bm{\theta}$ into two parts, i.e., $\bm{\theta}=(\bm{\theta}_1,\bm{\theta}_2)$. In particular,  we capture knowledge of the current state in a small-size latent probabilistic context variable $Z$ with $\bm{\theta}_1$, and use the latent context to inference the action with $\bm{\theta}_2$, as shown in Fig.~\ref{actor}.

An optimal representation of $S$
would capture the relevant factors and compress $S$ by diminishing the irrelevant parts which do not contribute to the prediction of action $A$. In a Markovian structure $S\to Z \to A$, $S$ is the input (state), $Z$ is the representation of state $S$, and $A$ is the label (optimal action) of $S$. Accordingly,  the information bottleneck principle seeks an embedding distribution $p_{{\bm{\theta}}_1^{*}}(\bm{z}|\bm{s})$ such that:
\begin{equation}
	\begin{aligned}
		p_{{\bm{\theta}}_1^{*}}(\bm{z}|\bm{s})&=\arg \underset{p_{{\bm{\theta}}_1}(\bm{z}|\bm{s})}{\max} I(A;Z)-\beta I(S;Z),\\
		&=\arg \underset{p_{{\bm{\theta}}_1}(\bm{z}|\bm{s})}{\max}H(A)-H(A|Z)-\beta I(S;Z),\\
		&=\arg \underset{p_{{\bm{\theta}}_1}(\bm{z}|\bm{s})}{\max}-H(A|Z)-\beta I(S;Z),\\
	\end{aligned}
\end{equation}
where $H(A|Z)=\int p(\bm{a},\bm{z})\log p(\bm{a}|\bm{z}) d\bm{a} d\bm{z}$ denotes the condition entropy, and $\beta>0$ controlling the tradeoff between the compression and prediction.  

Unfortunately, in video streaming system, the optimal bitrate decision is hard to obtain. Taking the future-aware optimization solution as the expert demonstration $\hat{A}$, another goal of the IB design is to prevent memorizing future network in the training data, which is attained by maximizing the mutual information between  the encoded feature $Z$ and the adversarial expert demonstration that may suboptimal but without any future knowledge $\tilde{A}$\footnote{Without loss of generality, we adopt the RobustMPC~\cite{10.1145/2785956.2787486} solution as the adversarial expert in our simulation, where the harmonic mean of past throughput is utilized to estimate the future throughput.}, i.e., $I(\tilde{A},Z)$. 
By combing the goals of relevant information extraction, irrelevant information forgetting, and future information adversary, we propose a new principle, future Adversarial Information Bottleneck (AIB), which is formulated by an optimization problem that maximizes the following objective function:
\begin{equation}
	\begin{aligned}
		&\underset{p_{\hat{\bm{\theta}}}(\bm{z}|\bm{s})}{\max}-H(\hat{A}|Z)-\beta I(S;Z)+\eta I(\tilde{A}|Z),\\
		=&\underset{p_{\hat{\bm{\theta}}}(\bm{z}|\bm{s})}{\max}-H(\hat{A}|Z)-\beta I(S;Z)+\eta[H(\tilde{A})-H(\tilde{A}|Z)],\\
		=&\underset{p_{\hat{\bm{\theta}}}(\bm{z}|\bm{s})}{\min}H(\underbrace{\hat{A}}_{\text{Expert}}|Z)+\beta I(S;Z)+\eta H(\underbrace{\tilde{A}}_{\text{Adverse Expert}}|Z),
	\end{aligned}
	\label{mutualinformationformulation}
\end{equation}
where $\eta>0$ also controls the tradeoff between prediction and future adversary.

\subsection{Variational Approximation of AIB Objective}
However, the distributions $p(\bm{z})$ and $p(\bm{a}|\bm{z})$ are generally intractable due to the high dimensionality. 
Similar to \cite{9837474}, we adopt tools from variational inference to approximate these intractable distributions. 
The idea of the variational approximation is to posit a family of distributions and find a member of that family which is close to the target distribution.
Specifically, 
we formulate the distributions $p_{\bm{\bm{\theta}}_1}(\bm{z}|\bm{s})$ 
and $p_{\bm{\bm{\theta}}_2}(\bm{a}|\bm{z})$ in terms of Deep Neural Networks (DNNs) 
with parameter $\bm{\bm{\theta}}_1$ and $\bm{\bm{\theta}}_2$. A common approach to parameterize 
the condition distribution  $p_{\bm{\bm{\theta}}_1}(\bm{z}|\bm{s})$ is adopting the multivariate 
Gaussian distribution, 
i.e., $p_{\bm{\bm{\theta}}_1}(\bm{z}|\bm{s})=\mathcal{N}(\bm{z}|\bm{\mu},\bm{\sigma})$, 
where the mean $\bm{\mu}$ and variance $\bm{\sigma}$ are the output of the DNN parameterized 
with ${\bm{\theta}_1}$. Besides, we approximate the prior distribution $p(\bm{z})$ with a centered 
isotropic Gaussian distribution $q(z)=\mathcal{N}(\bm{z}|\bm{0},\bm{e})$, where $\bm{e}$ denotes 
the {identity} matrix.
Accordingly, we recast the objective function in (\ref{mutualinformationformulation}) as follows:
\begin{equation}
	\begin{aligned}
		\mathcal{L}_{AIB}=\mathbb{E}_{p(\bm{s},\bm{a})}\Big[&
		\mathbb{E}_{p_{\bm{\bm{\theta}}_1}(\bm{z}|\bm{s})}\big[
	\log p_{\bm{\bm{\theta}}_2}(\hat{\bm{a}}|\bm{z})+\eta\log p_{\bm{\bm{\theta}}_2}(\tilde{\bm{a}}|\bm{z})\big]\\&
		+\beta D_{KL}(p_{\bm{\bm{\theta}}_1}(\bm{z}|\bm{s})||q(\bm{z}))\Big].
	\end{aligned}
\end{equation}

	\begin{figure*}[!b]
		\rule{\textwidth}{1pt}
		\begin{equation}\label{aib_der}
			\begin{aligned}
				&H(\hat{A}|Z)+\eta H(\tilde{A}|Z)+\beta I(S;Z)\\
				=&\int p(\hat{\bm{a}},\bm{z}) \log(p(\hat{\bm{a}}|\bm{z}))d\hat{\bm{a}}d\bm{z}+\eta \int p(\tilde{\bm{a}},\bm{z}) \log(p(\tilde{\bm{a}}|\bm{z}))d\tilde{\bm{a}}d\bm{z}
				+\beta \int p(\bm{z}|\bm{s}) p(\bm{s}) \log(\frac{p_{\bm{\bm{\theta}}_1}(\bm{z}|\bm{s})}{p(\bm{z})})d\bm{s}d\bm{z} \\
				=&\int p(\hat{\bm{a}},\bm{z}) \log(q_{\bm{\bm{\theta}}_2}(\hat{\bm{a}}|\bm{z}))d\hat{\bm{a}}d\bm{z}+\eta \int p(\tilde{\bm{a}},\bm{z}) \log(q_{\bm{\bm{\theta}}_2}(\tilde{\bm{a}}|\bm{z}))d\tilde{\bm{a}}d\bm{z}+\beta \int p(\bm{z}|\bm{s}) p(\bm{s}) \log(\frac{p_{\bm{\bm{\theta}}_1}(\bm{z}|\bm{s})}{q(\bm{z})})d\bm{s}d\bm{z}\\
				&-D_{KL}(p(\hat{\bm{a}}|\bm{z})||q_{\bm{\bm{\theta}}_2}(\hat{\bm{a}}|\bm{z}))-\eta D_{KL}(p(\tilde{\bm{a}}|\bm{z})||q_{\bm{\bm{\theta}}_2}(\tilde{\bm{a}}|\bm{z}))-\beta D_{KL}(p(\bm{z})||q(\bm{z}))\\
				\le  &\int p(\hat{\bm{a}},\bm{z}) \log(q_{\bm{\bm{\theta}}_2}(\hat{\bm{a}}|\bm{z}))d\hat{\bm{a}}d\bm{z}+\eta \int p(\tilde{\bm{a}},\bm{z}) \log(q_{\bm{\bm{\theta}}_2}(\tilde{\bm{a}}|\bm{z}))d\tilde{\bm{a}}d\bm{z}+\beta \int p(\bm{z}|\bm{s}) p(\bm{s}) \log(\frac{p_{\bm{\bm{\theta}}_1}(\bm{z}|\bm{s})}{q(\bm{z})})d\bm{s}d\bm{z} = \mathcal{L}_{AIB}.
			\end{aligned}
		\end{equation}
	\end{figure*}
The above formulation is termed as the variational approximation, which serves an upper bound on the AIB objective in (\ref{mutualinformationformulation}). The detailed derivations are  shown in (\ref{aib_der}), where KL-divergence is non-negative and thus eliminated in the bound. To optimize the objective (\ref{mutualinformationformulation}) using stochastic gradient descent,  we further apply the reparameterization trick \cite{9837474}, where the Monte Carlo estimation  in (\ref{mcmc}) is differentiable with respect to $\bm{\theta}$. In particular, given a minibatch of state-action tuple $\{s_i,a_i\}^{M}_{i=1}$, we have the following empirical estimation:
\begin{equation}\label{mcmc}
	\begin{aligned}
		&\mathcal{L}_{AIB}\simeq\frac{1}{M}\sum_{m=1}^{M}
		\Big[\\ &
		\log p_{\tilde{\bm{\theta}}}(\hat{\bm{a}}_m|\bm{z}_m)\eta\log p_{\tilde{\bm{\theta}}}(\tilde{\bm{a}}_m|\bm{z}_m)
		+\beta D_{KL}(p_{\hat{\bm{\theta}}}(\bm{z}|\bm{s}_m)||q(\bm{z}))\Big].
	\end{aligned}
\end{equation}
The overall training process is summarized in Algorithm~1.
\begin{algorithm}[!t]
	\caption{\small Training ABABR Agent.}\label{alg:cap}
	\begin{algorithmic}
		\Require Alternative Optimization Solver (Section 5.3)
		\While{EPOCH$\;= 1$ to MaxEP}
		\State Randomly pick a trace from the network dataset
		\State $i \gets 1$
		{
		\State Initialize the State $\bm{S}_1$
		\State Clear state-action set}
		\While{Not the end of the video}
		\State Get ABR State $\bm{S}_i$
		\State Pick $a_i$ according to policy $\pi_{\theta}(s_i)$
		\State Compute the expert $\hat{a}_i$ and $\tilde{a}_i$ with Solver
		{
		\State Combine $(s_i,\hat{a}_i,\tilde{a}_i)$ into state-action set}
		\State Sample a minibatch from the state-action set
		\State Update network $\theta$ with the minibatch using Eq.~(\ref{mcmc})
		\State Execute action $a_i$, $i \gets i+1$
		\EndWhile
		\EndWhile
	\end{algorithmic}
\end{algorithm}
    \begin{table}[!t]
	\caption{\small Details of Video Datasets and QoE Parameters. The experimental setup is referenced to \cite{10.1145/3452296.3472923} and \cite{10.1145/3098822.3098843}.}
	\begin{center}
			\resizebox{\linewidth}{!}{
				\begin{tabular}{ccccccc}
					\toprule
					\textbf{Video} & \makecell[c]{\textbf{Bitrate Levels} \textbf{(Mbps)}}
					& $L$
					& $I$ & $\alpha_1$ & $\alpha_2$ & \textbf{RTT} \\
					\midrule
					{
						Provided by $\text{A}^2$BR} & \{20, 40, 60, 80, 110, 160\} & 4s & 39 & 160 & 1  & 104ms \\
					{
						Provided by Pensieve} & \{0.3, 0.75, 1.2, 1.85, 2.85, 4.3\} & 4s & 48 & 4.3 & 1  & 80ms \\
					\bottomrule
			\end{tabular}}
			\label{tab1}
		\end{center}
	\end{table}
\section{Simulations}
{
In order to evaluate the performance of ABABR,
we simulate the adaptive video streaming process by using the real-world network throughput
datasets (HSDPA\cite{riiser2013commute}/FCC16 \cite{fcc}, and Lumos5G \cite{narayanan2020lumos5g}).
Each dataset is divided into a training set and a testing set. 
We use the training dataset to train the actor network. 
All experimental results below are the average results under five random seeds on the testing dataset.}
Without loss of generality, we utilize the linear QoE metric, where $q(r_i)=r_i$.  
Details of the video datasets and the corresponding QoE parameters are shown in Table~\ref{tab1}. 
We compare ABABR with five representative ABR methods:
\begin{itemize}
	\item DASH: the conventional heuristic-based ABR methods. In our experiment, we employ the buffer-based  heuristic based solely on buffer occupancy, i.e., bandwidth-DASH. 
	\item RobustMPC~\cite{10.1145/2785956.2787486}: QoE metrics are maximized using the 
	MPC framework by observing the dynamics of buffer occupancy and throughput.
	\item Pensieve~\cite{10.1145/3098822.3098843}: an RL-based algorithm which takes the
	former network status as states and optimizes itself with
	various network conditions using A3C method. 
	\item Comyco~\cite{10.1145/3343031.3351014}: a behavior cloning imitation learning algorithm. In our experiment, comyco is trained with the first term of eq.~\ref{mutualinformationformulation} and the same expert demonstrations as ABABR. 
	\item MERINA~\cite{10.1145/3503161.3548331}: a Meta-RL based algorithm which consists of a probabilistic latent encoder and meta-policy network. Additionally, imitation learning-based pre-training is leveraged
	as a meta-policy search scheme. \footnote{The proposed ABABR is orthogonal to the meta-learning approach, and we can also use the meta-learning approach to further improve performance. In our experiments, We employ MERINA as a state-of-the-art benchmark to evaluate the performance of the proposed method.
	We show that the proposed method outperforms this training methods in a variety of networks, although not enhanced with meta-learning training techniques.} 
\end{itemize}
	For fair comparison, we retrained Pensieve, Comyco, and MERINA with the same neural network, videos, datasets, and QoE metrics.

    	\begin{table*}[!t]
    		\caption{\small Performance details of different ABR algorithms under five random seeds.}
    		\centering
    		\begin{tabular}{c|ccccc|ccccc}
    			\toprule
    			\multirow{3}{*}{Alg.} & \multicolumn{5}{c|}{3G} & \multicolumn{5}{c}{5G}    \\ 
    			\cmidrule{2-11}
    			
    			& \multicolumn{2}{c|}{QoE} & \multicolumn{1}{c|}{\multirow{2}{*}{Avg Bit}}
    			& \multicolumn{1}{c|}{\multirow{2}{*}{Avg Reb}} & \multirow{2}{*}{Avg Var} 
    			& \multicolumn{2}{c|}{QoE} & \multicolumn{1}{c|}{\multirow{2}{*}{Avg Bit}} 
    			& \multicolumn{1}{c|}{\multirow{2}{*}{Avg Reb}} & \multirow{2}{*}{Avg Var} \\
    			\cmidrule{2-3} \cmidrule{7-8}
    			
    			& \multicolumn{1}{c|}{Avg} & \multicolumn{1}{c|}{Max/Min} & \multicolumn{1}{c|}{} & \multicolumn{1}{c|}{} & \multicolumn{1}{c|}{} & \multicolumn{1}{c|}{Avg} & \multicolumn{1}{c|}{Max/Min} & \multicolumn{1}{c|}{} & \multicolumn{1}{c|}{} & \multicolumn{1}{c}{}  \\
    			
    			\midrule
    			DASH      & 0.312 & +0/-0 & 1.037 & 0.365 & 0.368 & 
    			104.060 & +0/-0 & 120.940 & 7.336 & 9.795 \\
    			RobustMPC & 0.495 & +0.002/-0.001 & 1.149 & 0.511 & 0.146 & 
    			103.189 & +0.736/-0.186 & 126.288 & 12.547 & 10.830 \\
    			Pensieve  & 0.522 & +0.023/-0.028 & 1.038 & 0.401 & 0.115 &
    			90.037 & +23.116/-29.801 & 109.091 & 13.379 & 5.825 \\
    			Comyco    & \underline{0.548} & +0.026/-0.030 & 1.065 & 0.432 & 0.080 &
    			105.621 & +2.720/-1.223 & 141.077 & 26.755 & 8.930 \\
    			MERINA    & 0.538 & +0.010/-0.016 & 1.133 & 0.450 & 0.151 &
    			\underline{107.368} & +2.968/-2.606 & 140.291 & 20.581 & 12.667 \\
    			\midrule
    			\textbf{ABABR}     &  \textbf{0.588} & +0.011/-0.008 & 1.131 & 0.436 & 0.097 &
    			\textbf{110.284} & +1.154/-2.516 & 140.242 & 21.577 & 8.602 \\ 
    			\bottomrule  
    		\end{tabular} \label{table:details}
    	\end{table*}
		Without loss of generality, we configure the actor network of ABABR, Comyco, and MERINA with the same structure and hyper-parameters as Pensieve \cite{10.1145/3098822.3098843}. 
		In particular, the 1D-CNN layers are set as 128 filters each of size 4 with stride 1, 
		and the fully connected layers use 128 neurons.
		We use relu as activation function and Adam as optimizer. 
          The learning rates is configured to  $10^{-4}$. 
          Similar to \cite{10.1145/3098822.3098843}, the maximum buffer occupancy is set as 60 seconds 
          and the last $k=8$ state are fed into the network. Unless otherwise specified, we set $N$, $\beta$, and $\eta$ as 8, 0.0001, and 0.2, respectively. 


\begin{figure}[!t]
	\centering
	\begin{subfigure}{.99\linewidth}
		\centering
		\includegraphics[clip,  trim={0.2cm 0cm 0.8cm 0cm}, width=\linewidth]{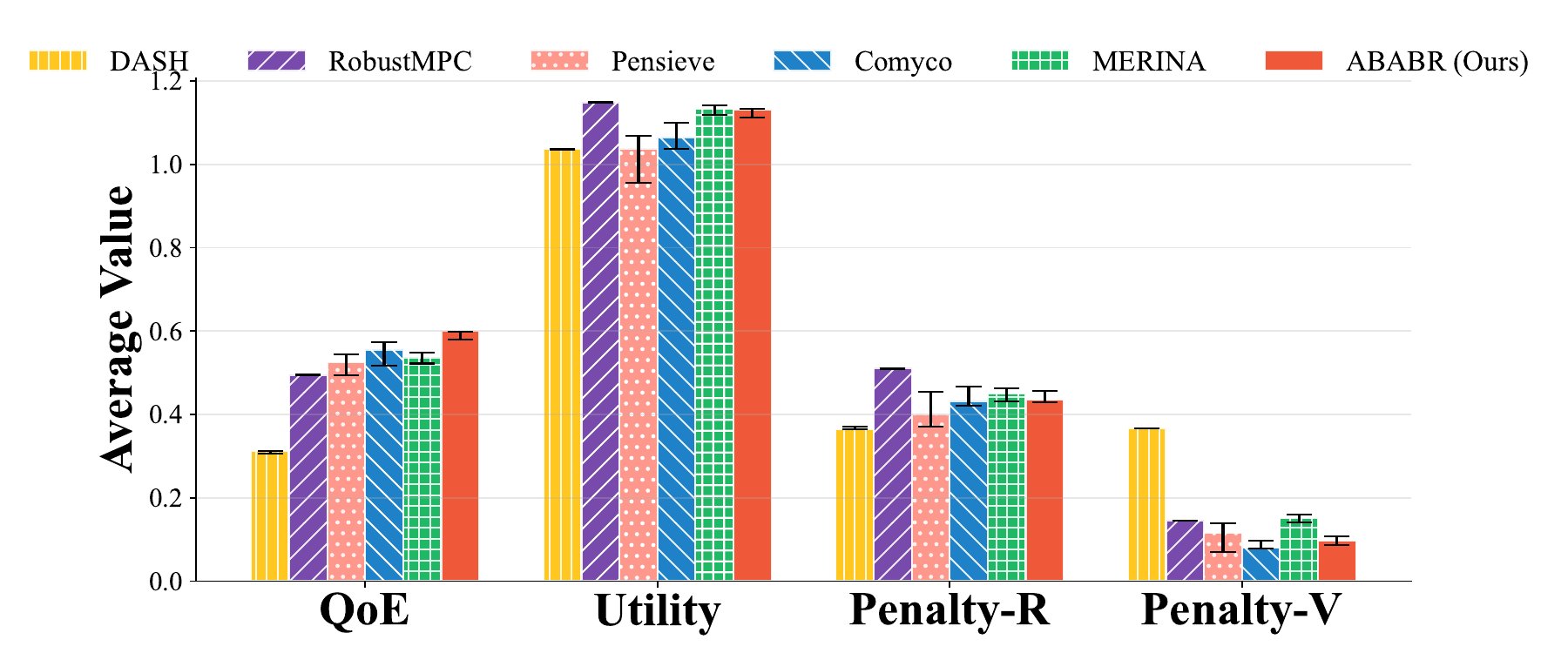}
		\caption{\small QoE in HSDPA/FCC16}
	\end{subfigure}
	~
		\begin{subfigure}{.99\linewidth}
	\centering
	\includegraphics[clip,  trim={0.2cm 0cm 0.8cm 0cm},width=\linewidth]{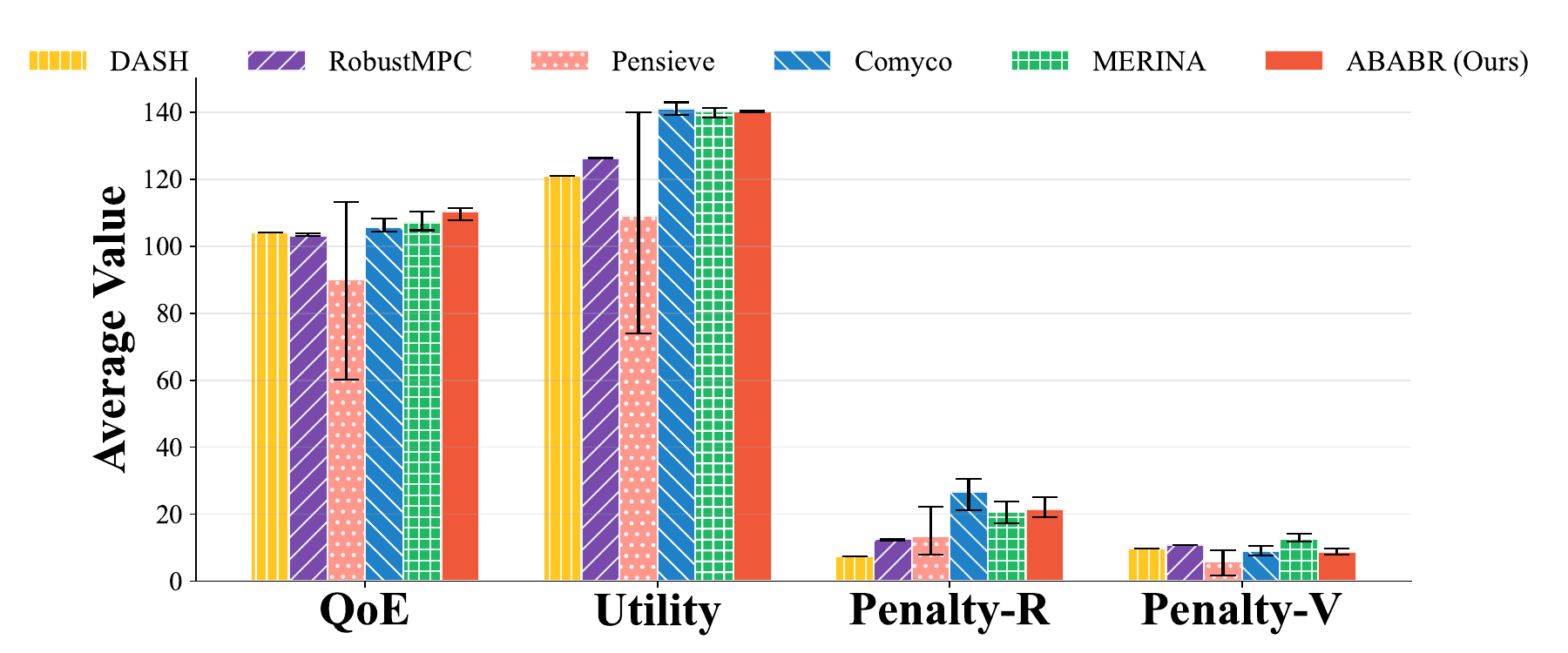}
	\caption{\small QoE in Lumos5G}
	\end{subfigure}
	\caption{\small Performance comparison  versus  ABR algorithms.}\label{fig:Performance} 
\end{figure}

          We evaluate ABABR to answer the following questions:
          \begin{itemize}
            \item Does the ABABR  improve  both trace-average QoE and trace-wise ranking  in various cellular networks? (Section~6.1)
            \item Does the ABABR  successfully improve the sample efficiency and computation overhead? (Section~6.2)
            \item How does each component of ABABR contribute to the performance gain? (Section 6.3)
          \end{itemize}
\subsection{Performance versus Various Networks}

\subsubsection{Average QoE versus Various Networks}
\begin{figure}[!t]
	\begin{subfigure}{.48\linewidth}
		\centering
		\includegraphics[clip,  trim={0.1cm 0cm 1cm 0.5cm},width=\linewidth]{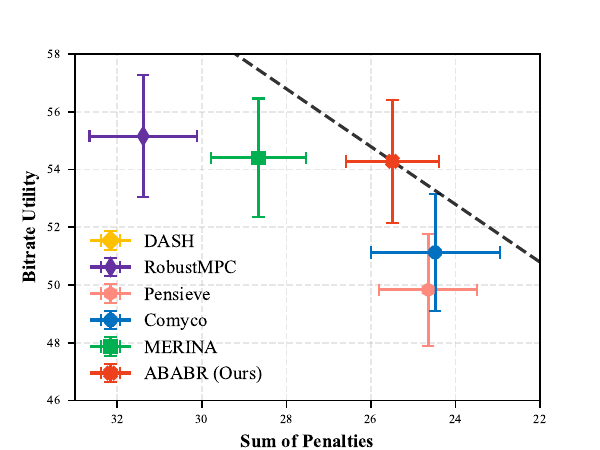}
		\caption{\small  HSDPA/FCC16}
	\end{subfigure}
	~
	\begin{subfigure}{.48\linewidth}
		\centering
		\includegraphics[clip,  trim={0.1cm 0cm 1cm 0.5cm},width=\linewidth]{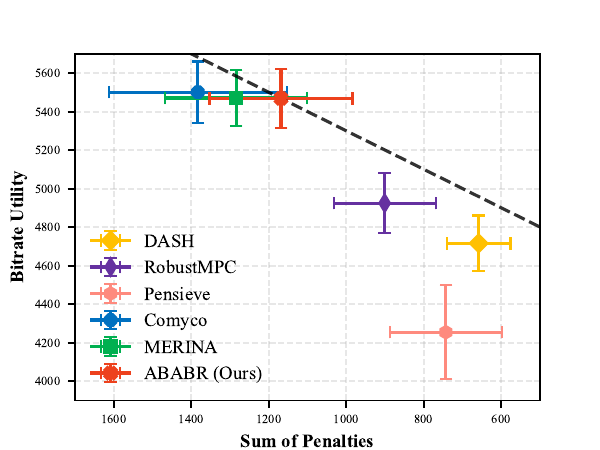}
		\caption{\small Lumos5G}
	\end{subfigure}
	\caption{\small Trade-offs comparison versus ABR algorithms.}\label{fig:ErrorBar} 
\end{figure}

In the first experiment, we assess the performance of ABABR on the both testing datasets. 
The comparison results, illustrated in Fig.~\ref{fig:Performance} and Table~\ref{table:details}, reveal the average Quality of Experience (QoE) and its components with remarkable specificity. 
Notably, ABABR achieves astounding results, surpassing other representative techniques. In the HSDPA/FCC16 dataset, it outperforms DASH, RobustMPC, Pensieve, Comyco, and MERINA by $88.46$\%, $18.78$\%, $12.64$\%, $7.30$\%, and $9.29$\% higher average QoE, respectively. 
Specifically, ABABR reduces the rebuffer and variation penalties as much as possible in the case of selecting a high bitrate. 
For instance, Fig.~\ref{fig:ErrorBar}a distinctly illustrates our algorithm's superior ability to strike a more optimal balance among the three QoE factors, resulting in an overall higher QoE.
Compared with RobustMPC and MERINA, ABABR can reduce the sum of penalties by $18.79$\% and $11.06$\% when the bitrate utility is similar. What's more, ABABR increases the bitrate utility by $8.93$\% and $6.17$\% compared with Pensieve and Comyco while keeping the sum of penalties similar. 
Similarly, In Fig.~\ref{fig:Performance}b and~\ref{fig:ErrorBar}b, the superiority of ABABR can also be extended to 5G scenarios, where ABABR outperforms all benchmark methods. 
This extraordinary performance is attributed to \emph{the proposed adversarial information bottleneck method, which effectively filters out task-irrelevant information from the encoded feature vector, consequently leading to a more robust performance on unseen network traces}. 
From Fig.~\ref{fig:Performance} and Table~\ref{table:details}, we observe that ABABR consistently 
achieves two of the high bitrates in both the two datasets, while simultaneously maintaining 
low variation and rebuffer penalties. 
In contrast, all the benchmarks exhibit varying performance in these two networks. 
This remarkable characteristic sets ABABR apart, as it not only provides the highest QoE performance but also ensures stable and reliable performance across various networks.

\subsubsection{Rank Scores versus Various Algorithms}
\begin{table}[!t]
	\caption{\small Ranking statistics of each traces in HSDPA testing dataset. AR: Average Ranking.}
	\centering
	\resizebox{\linewidth}{!}{
	\begin{tabular}{c|llllll|ll}
		\toprule
		Alg.      & R1(\%) & R2 & R3 & R4 & R5 & R6  & AR  & Point\\
		\midrule
		DASH      & 00.7   & 00.7   & 03.5   & 04.9	  & 14.1   & 76.1 & 5.6  & 1266    \\
		R-MPC & 14.0  & 26.1  & 12.7  & 16.9  & 19.7 & 10.6   & 3.3 &  2124   \\
		Pensieve  & 04.2   & 09.2  & 19.7  & 23.2  & 43.0 & 00.7    & 3.9  & 1818    \\
		Comyco    & \underline{19.0}  & 23.2  & 26.1  & 17.6 &  08.5 &	 05.6	  & \underline{2.9}  & \underline{2308}    \\
		MERINA    & 12.0  & 17.6  & 26.8  & 28.9 & 12.0  & 02.8   & 3.2 &  2139   \\
		\midrule
		\textbf{ABABR}     & \textbf{50.0}  & 23.2  & 11.3  & 08.5  & 02.8 &	04.2	  &  \textbf{2.0} &   \textbf{2841}  \\
		\bottomrule  
	\end{tabular} }
	\label{table:rank}
\end{table}
The QoE function, in reality, represents a score rather than an absolute value. Its physical significance is not tied to a specific numerical value but rather to its relative positioning. That is, a higher QoE for an individual video session can indeed indicate that the user experiences greater satisfaction during that specific session. However, having a higher average QoE across various video sessions does not necessarily translate to an overall higher satisfaction \cite{10.1145/3386290.3396930}. Averaging QoE scores may mask variations in satisfaction levels across different sessions. 

To underscore the superiority of our algorithm, we computed the ranking statistics for each trace in the HSDPA testing dataset, as presented in Table.~\ref{table:rank}\footnote{Without loss of generality, we compute the scores with the current point system of Formula One \url{https://fantasy.formula1.com/en/game-rules}. That is $25$, $18$, $15$, $12$, $10$, and $8$ points for 1st, 2nd, 3rd, 4th, 5th, and 6th places, respectively.}. A comparison with Comyco reveals that ABABR exhibits a $23.09$\% point improvement and a $30.01$\% reduction in the average ranking. In comparison with the meta RL benchmark MERINA, ABABR achieves an additional $32.82$\% points and lowers the average ranking by $37.50$\%. Notably, ABABR attains the best QoE in more than $50$\% of traces, surpassing RL-based benchmark MERINA and Pensieve, which only excel in $12$\% and $4.2$\% of traces, respectively. This underscores ABABR's robustness as an ABR policy, \emph{consistently delivering best results}.

Moreover, we can also observe from Table.~\ref{table:rank} that  RL-based benchmarks often achieve Rank-3 to Rank-5 performances. We posit that RL-based methods successfully avoid very low QoE cases (bad cases) due to extensive exploration and interaction with the environment, thereby improving average QoE. However, they fall short of achieving the best performance in our experiment, possibly because the complex optimal solution of MINLP is challenging to search and explore. 

On the other hand, IL-based methods, such as ABABR and Comyco, secure most of the Rank-1 performances, thanks to supervision from the offline optimal. This, coupled with the relative nature of QoE scores, further validates the superiority of both imitation learning approaches and the proposed ABABR algorithm.

\begin{figure*}[!t]
	\centering
	\begin{subfigure}{.321\linewidth}
		\centering
		\includegraphics[clip,  trim={0.6cm 0.4cm 1.5cm 0.5cm}, width=\linewidth]{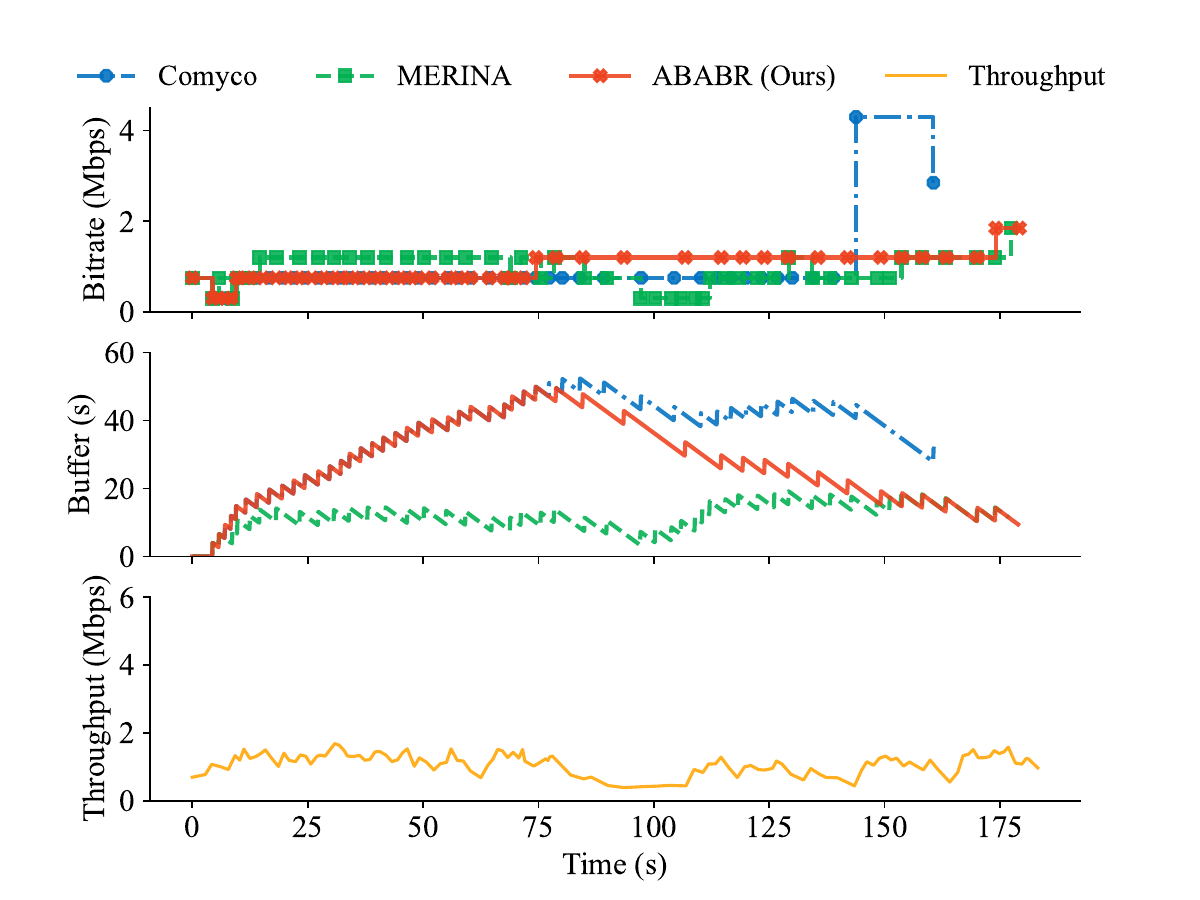}
		\caption{\small Slow-network path.}
	\end{subfigure}
	~
	\begin{subfigure}{.321\linewidth}
		\centering
		\includegraphics[clip,  trim={0.6cm 0.4cm 1.5cm 0.5cm},width=\linewidth]{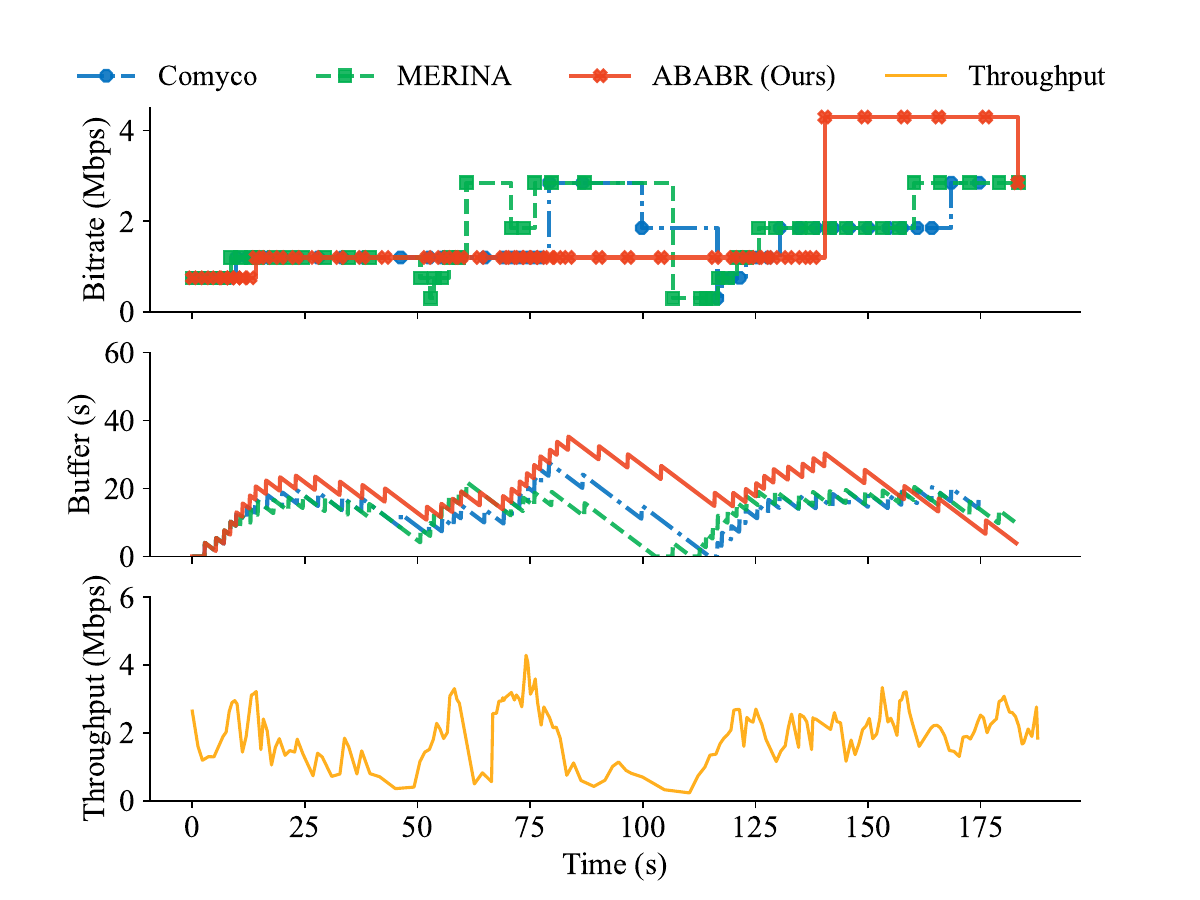}
		\caption{\small Medium-network path.}
	\end{subfigure}
	~
	\begin{subfigure}{.321\linewidth}
		\centering
		\includegraphics[clip,  trim={0.6cm 0.4cm 1.5cm 0.5cm},width=\linewidth]{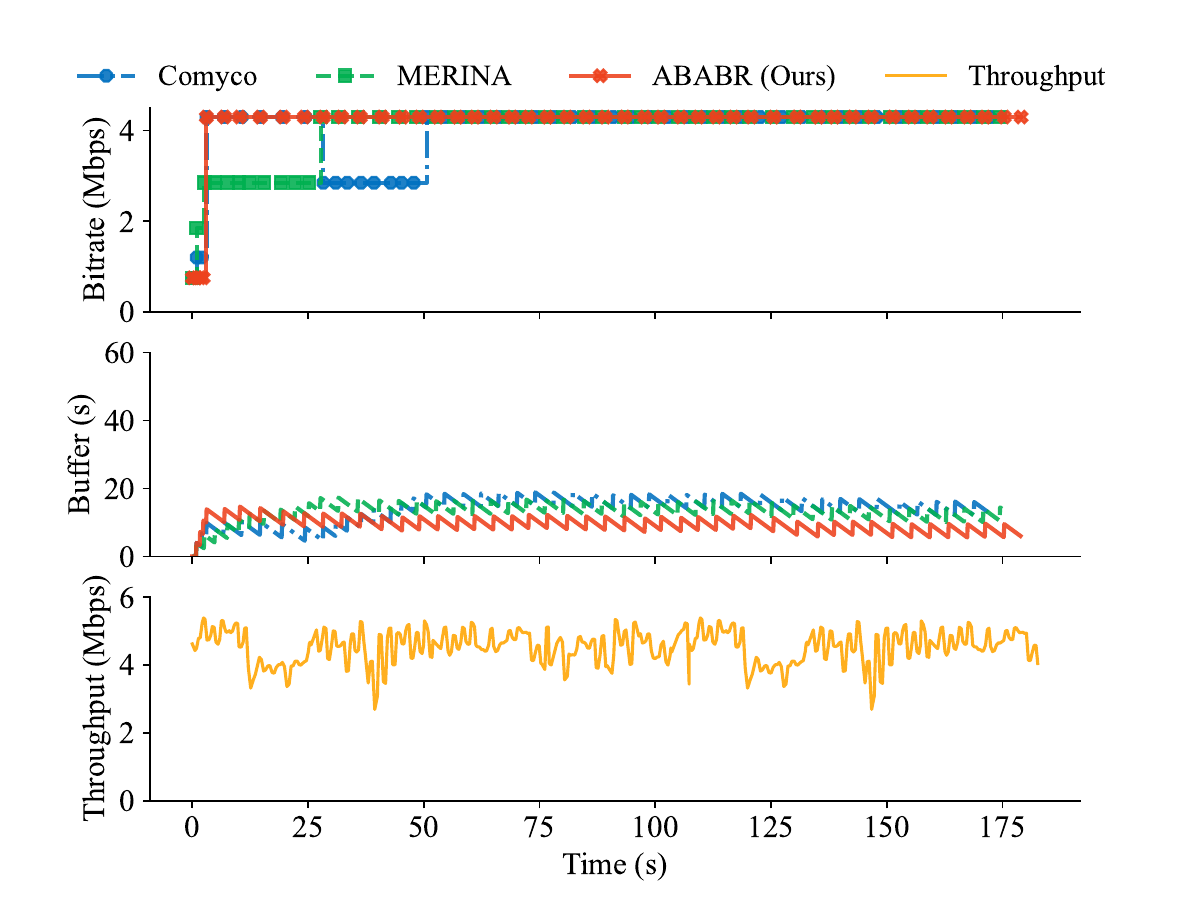}
		\caption{\small Fast-network path.}
	\end{subfigure}
	\caption{\small Policy comparison between different ABR algorithms in different types of network conditions.}
	\label{fig:casestudy} 
	
\end{figure*}

\subsubsection{Case Study} We further compare the learned policies by ABABR with two representative benchmarks, Comyco and MERINA. 
In Fig.~\ref{fig:casestudy}, we plot the status information generated by different  
policies under three different network conditions during the video session.

In the slow- and medium-network path (Fig.~\ref{fig:casestudy}a-b), ABABR and Comyco both start by choosing video chunks with smaller bitrates to ensure an adequate buffer size initially. However, when the buffer size reaches a certain threshold, they attempt to increase the bitrate despite the suboptimal network state. Alternatively, MERINA frequently changes the bitrate to handle network fluctuations, incurring a high fluctuation penalty. 
In the fast-network (Fig.~\ref{fig:casestudy}c), all methods initially choose a relatively low bitrate and switch to a higher bitrate later. ABABR selects a low bitrate only in the first two chunks, maintaining the highest bitrate for the rest of the video session, while the benchmarks switch to a high bitrate much later. 

To sum up, Fig.~\ref{fig:casestudy} demonstrates a preference for selecting video chunks with smaller bitrates in the early stages to prioritize building a sufficient buffer size.  As the buffer size reaches a threshold, adjustments are made based on considerations of buffer size and prevailing network conditions. However, the behavior cloning IL-based method Comyco tends to make suboptimal decisions on when to switch to a higher bitrate due to its limited generalization ability. ABABR, with its adversarial information bottleneck framework, achieves better performance. In contrast, the meta-RL-based approach MERINA pays little attention to buffer size accumulation, making it more susceptible to network fluctuations.

\begin{figure}[!t]
	\centering
	\begin{subfigure}{.48\linewidth}
		\centering
		\includegraphics[clip,  trim={0.4cm 0cm 1.5cm 1.5cm},width=\linewidth]{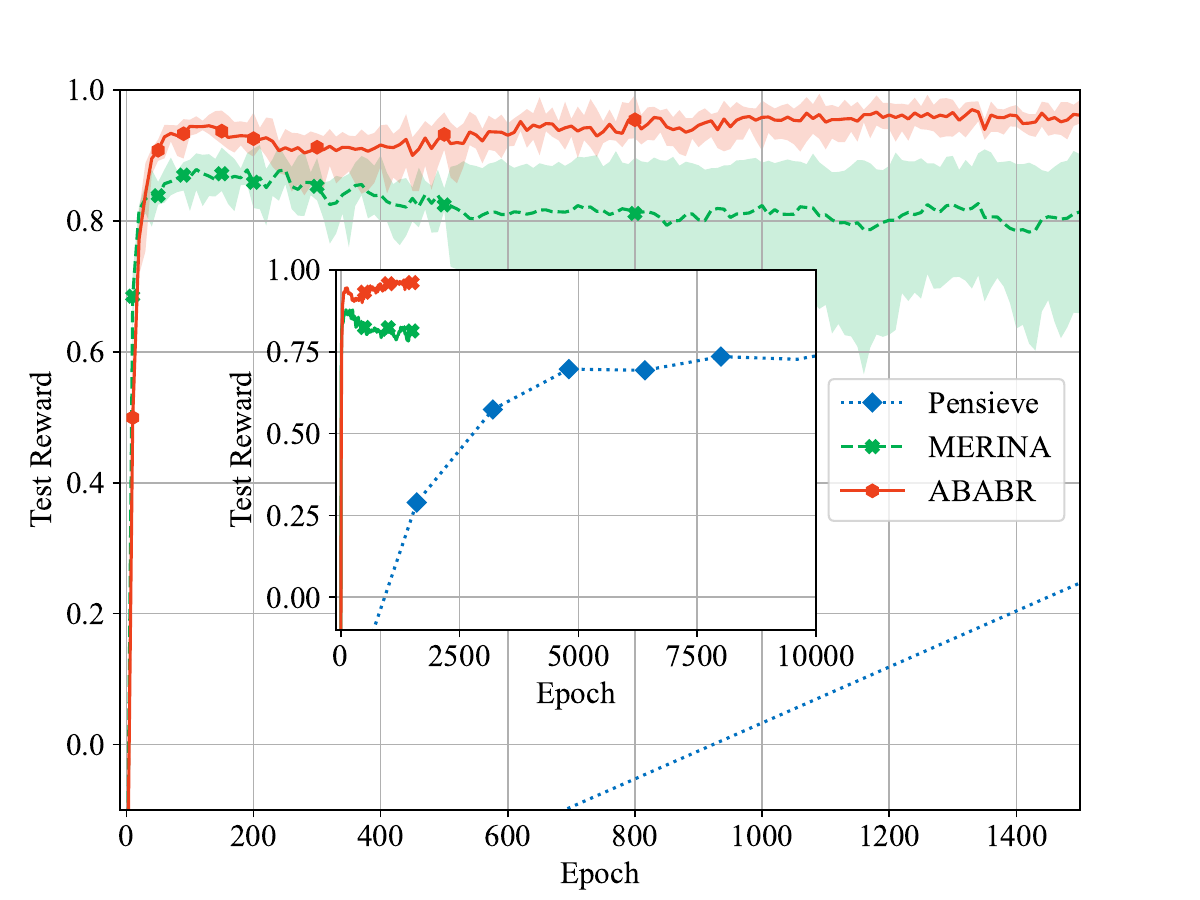} 
		\caption{\small Convergence}
	\end{subfigure}
	~
	\begin{subfigure}{.48\linewidth}
		\centering
		\includegraphics[clip,  trim={0.4cm 0cm 1.5cm 1.5cm},width=\linewidth]{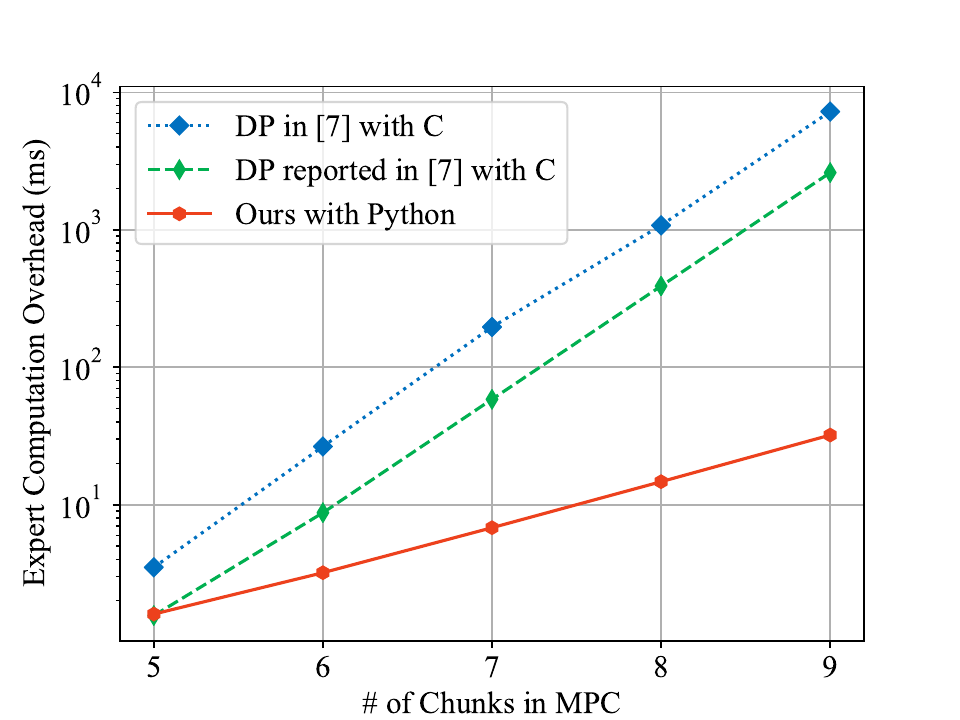}
		\caption{\small Computation Overhead}
	\end{subfigure}
	\caption{\small Training overhead comparison. DP: Dynamic Programming.}
	\label{fig:compute} 
\end{figure}

\subsection{Convergence and Complexity Comparison}
In the second experiment, we plot the training convergence of the proposed ABABR, Pensieve and MERINA, as depicted in Fig.~\ref{fig:compute}a.
In particular, MERINA trains the actor network by imitation the RobustMPC expert during the first 500 epochs, and then update the actor and critic networks in a  meta RL manner.  
 Here, we observe that both ABABR and MERINA converge after about $100$ epochs, showcasing the remarkable efficiency of imitation learning. In contrast, RL-based methods suffer from significantly lower convergence rates. The reason behind this  difference lies in the agent's requirement to simultaneously learn a latent representation and a control policy, making generalization via a continuous reward signal inefficient and prone to suboptimal convergence. However, ABABR overcomes these challenges by leveraging prior knowledge from expert demonstrations and applying the information bottleneck principle, enabling easy and effective learning of the latent space representation without overfitting to the dataset.


In the third experiment, we delve into the computation time for expert demonstrations, comparing the proposed Alternative Optimization (AO) method with the conventional Dynamic Programming (DP) method from \cite{10.1145/3343031.3351014}, 
with varying future horizon lengths ($N$). 
In particular, we implement the alternative optimization via CVXPY with MOSEK solver on Python, and the dynamic programming method in C. 
As illustrated in Fig.~\ref{fig:compute}b, the proposed AO method drastically improves the efficiency of solving the MINLP for expert demonstration, boasting an average speed that is $146.3\times$  faster than the DP method. This remarkable advantage widens further with the increase in $N$, facilitating large-scale training and, in turn, improving the performance of imitation learning in video streaming.

\subsection{Adversarial Information Bottleneck Principle}
	\begin{figure}
	\centering
	\includegraphics[clip, trim={0.4cm 0cm 1.5cm 1.5cm},width=0.75\linewidth]{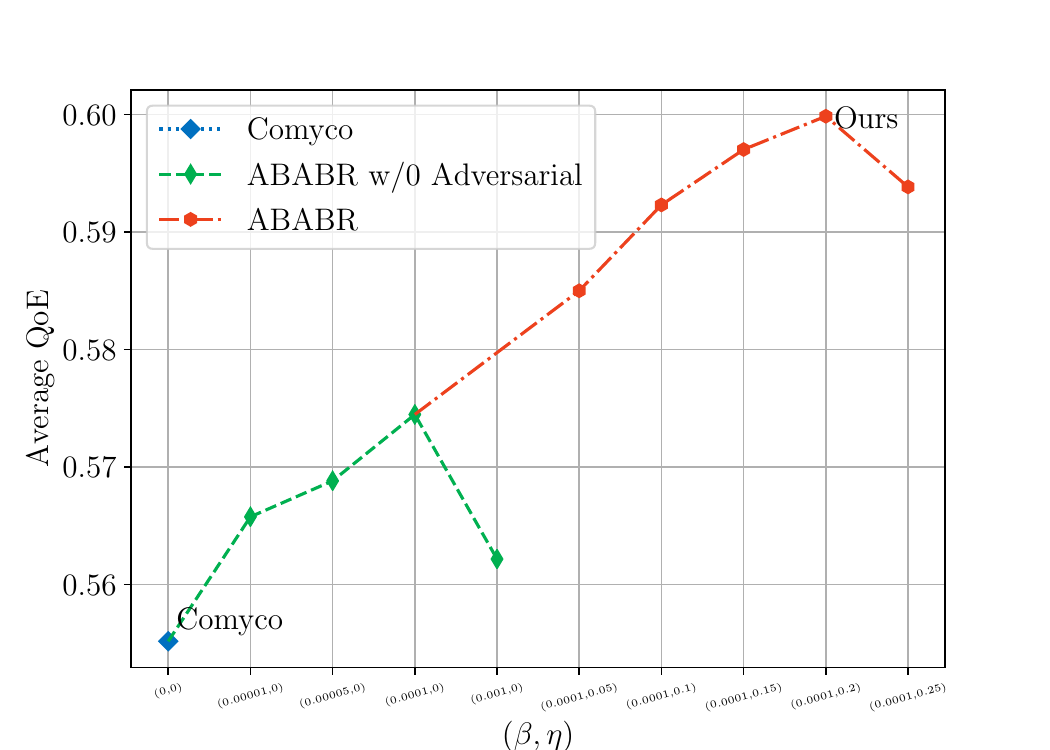}
	\caption{\small Ablation Study.}\label{AIB} 
\end{figure}
In the forth experiment, we meticulously evaluate the benefits of the task-relevant feature extraction (Information Bottleneck Principle) and the future adversarial principle, as demonstrated in Fig.~\ref{AIB}. 
By plotting the average QoE in LTE against different compression-prediction weight ($\beta$) and prediction-adversarial weight ($\eta$), we obtain valuable insights. The results show that increasing $\beta$  from $0$ to $0.0001$ enhances average QoE, validating the compression-prediction tradeoff's ability to improve the representation capacity of the latent space and consequently enhance performance on test traces. Similarly, augmenting $\eta$ from $0$ to $0.2$ also leads to increased average QoE. Overall, the AIB principle efficiently discards both task-irrelevant information and leaked future information, enabling a robust and generalizable latent space representation.


\section{Conclusions}
In this paper, we have presented a novel approach that combines imitation learning with the information bottleneck technique to improve learning-based adaptive video streaming algorithms. 
By leveraging imitation learning, alternative optimization, and the information bottleneck framework, we have achieved superior video quality in terms of both trace-average QoE and trace-wise ranking,  while mitigating issues such as overfitting and performance degradation. The convergence time and computation overhead are both significantly reduced. 

Moreover, our work highlights the importance of optimization theory as an indispensable framework for network optimizations. The concept of using optimization formulation as a digital twin has gained recognition as a promising solution. However, the leakage of future information in this digital twin-aided framework has been largely ignored. Our proposed framework addresses this limitation and provides a more comprehensive solution.

\bibliographystyle{IEEEtran}
\bibliography{IEEEfull.bib}

\begin{thebibliography}{10}
\providecommand{\url}[1]{#1}
\csname url@samestyle\endcsname
\providecommand{\newblock}{\relax}
\providecommand{\bibinfo}[2]{#2}
\providecommand{\BIBentrySTDinterwordspacing}{\spaceskip=0pt\relax}
\providecommand{\BIBentryALTinterwordstretchfactor}{4}
\providecommand{\BIBentryALTinterwordspacing}{\spaceskip=\fontdimen2\font plus
\BIBentryALTinterwordstretchfactor\fontdimen3\font minus
  \fontdimen4\font\relax}
\providecommand{\BIBforeignlanguage}[2]{{%
\expandafter\ifx\csname l@#1\endcsname\relax
\typeout{** WARNING: IEEEtran.bst: No hyphenation pattern has been}%
\typeout{** loaded for the language `#1'. Using the pattern for}%
\typeout{** the default language instead.}%
\else
\language=\csname l@#1\endcsname
\fi
#2}}
\providecommand{\BIBdecl}{\relax}
\BIBdecl

\bibitem{10.1145/2934872.2934898}
Y.~Sun,  \emph{et~al.}, ``Cs2p: Improving video bitrate selection and
  adaptation with data-driven throughput prediction,'' in \emph{Proc. Annu.
  Conf. ACM Spec. Interest Group Data Commun. Appl., Technol., Archit., Protoc.
  Comput. Commun. (SIGCOMM)}, 2016.

\bibitem{7524428}
K.~Spiteri \emph{et~al.}, ``Bola: Near-optimal bitrate adaptation for online
  videos,'' in \emph{IEEE Conference on Computer Communications (INFOCOM)},
  2016.

\bibitem{10.1145/2785956.2787486}
X.~Yin \emph{et~al.}, ``A control-theoretic approach for dynamic adaptive video
  streaming over http,'' in \emph{Proc. Annu. Conf. ACM Spec. Interest Group
  Data Commun. Appl., Technol., Archit., Protoc. Comput. Commun. (SIGCOMM)},
  2015.

\bibitem{10.1145/3098822.3098843}
H.~Mao \emph{et~al.}, ``Neural adaptive video streaming with pensieve,'' in
  \emph{Proc. Annu. Conf. ACM Spec. Interest Group Data Commun. Appl.,
  Technol., Archit., Protoc. Comput. Commun. (SIGCOMM)}, 2017.

\bibitem{8526814}
S.~Sengupta \emph{et~al.}, ``Hotdash: Hotspot aware adaptive video streaming
  using deep reinforcement learning,'' in \emph{Proc. Int. Conf. Netw. Protoc.
  (ICNP)}, 2018.

\bibitem{8737418}
B.~Alt, T.~Ballard, R.~Steinmetz, H.~Koeppl, and A.~Rizk, ``Cba: Contextual
  quality adaptation for adaptive bitrate video streaming,'' in \emph{IEEE
  Conference on Computer Communications (INFOCOM)}, 2019, pp. 1000--1008.

\bibitem{10.1145/3386290.3396930}
T.~Huang, R.-X. Zhang, and L.~Sun, ``Zwei: A self-play reinforcement learning
  framework for video transmission services,'' \emph{{IEEE} Trans. Multimedia},
  vol.~24, pp. 1350--1365, 2022.

\bibitem{9334431}
Z.~Meng \emph{et~al.}, ``Practically deploying heavyweight adaptive bitrate
  algorithms with teacher-student learning,'' \emph{{IEEE/ACM} Trans. Netw.},
  vol.~29, no.~2, pp. 723--736, 2021.

\bibitem{10.1145/3343031.3351014}
T.~Huang, C.~Zhou, X.~Yao, R.-X. Zhang, C.~Wu, B.~Yu, and L.~Sun,
  ``Quality-aware neural adaptive video streaming with lifelong imitation
  learning,'' \emph{{IEEE} J. Sel. Areas Commun.}, vol.~38, no.~10, pp.
  2324--2342, 2020.

\bibitem{9699071}
W.~Li, J.~Huang, S.~Wang, C.~Wu, S.~Liu, and J.~Wang, ``An apprenticeship
  learning approach for adaptive video streaming based on chunk quality and
  user preference,'' \emph{{IEEE} Trans. Multimedia}, vol.~25, pp. 2488--2502,
  2023.

\bibitem{9762785}
J.~Liu, Z.~Liu, J.~Huang, W.~Jiang, and J.~Wang, ``A buffer-based adaptive
  bitrate approach in wireless networks with iterative correction,'' \emph{IEEE
  Wireless Commun. Lett.}, vol.~11, no.~8, pp. 1644--1648, 2022.

\bibitem{9312489}
B.~Wang, M.~Xu, F.~Ren, C.~Zhou, and J.~Wu, ``Cratus: A lightweight and robust
  approach for mobile live streaming,'' \emph{{IEEE} Trans. Mobile Comput.},
  vol.~21, no.~8, pp. 2761--2775, 2022.

\bibitem{10.1145/3230543.3230558}
Z.~Akhtar,  \emph{et~al.}, ``Oboe: Auto-tuning video abr algorithms to network
  conditions,'' in \emph{Proc. Annu. Conf. ACM Spec. Interest Group Data
  Commun. Appl., Technol., Archit., Protoc. Comput. Commun. (SIGCOMM)}, 2018,
  p. 44–58.

\bibitem{9796948}
G.~Lv, Q.~Wu, Q.~Tan, W.~Wang, Z.~Li, and G.~Xie, ``Accurate throughput
  prediction for improving qoe in mobile adaptive streaming,'' \emph{{IEEE}
  Trans. Mobile Comput.}, pp. 1--18, 2023.

\bibitem{10041952}
Y.~Li, X.~Zhang, C.~Cui, S.~Wang, and S.~Ma, ``Fleet: Improving quality of
  experience for low-latency live video streaming,'' \emph{IEEE Trans Circuits
  Syst Video Technol}, pp. 1--1, 2023.

\bibitem{yan2020learning}
F.~Y. Yan, H.~Ayers, C.~Zhu, S.~Fouladi, J.~Hong, K.~Zhang, P.~Levis, and
  K.~Winstein, ``Learning in situ: a randomized experiment in video
  streaming,'' in \emph{Proc. USENIX Symp. Networked Syst. Des. Implement.
  (NSDI)}, 2020.

\bibitem{9860000}
J.~Pei, C.~An, A.~Zhou, L.~Liu, and H.~Ma, ``Par: Improving video bitrate
  adaptation via payload-aware throughput prediction,'' in \emph{IEEE
  International Conference on Multimedia and Expo (ICME)}, 2022.

\bibitem{kan2021uncertainty}
N.~Kan \emph{et~al.}, ``Uncertainty-aware robust adaptive video streaming with
  bayesian neural network and model predictive control,'' in \emph{NOSSDAV -
  Proc. Workshop Netw. Oper. Syst. Support Digit. Audio Video, Part MMSys},
  2021.

\bibitem{ICME23}
J.~Lin and S.~Wang, ``Improving robustness of learning-based adaptive video
  streaming in wildly fluctuating networks,'' in \emph{IEEE International
  Conference on Multimedia and Expo (ICME)}, 2023, pp. 1787--1792.

\bibitem{tianchihuang}
T.~Huang, C.~Zhou, R.-X. Zhang, C.~Wu, and L.~Sun, ``Buffer awareness neural
  adaptive video streaming for avoiding extra buffer consumption,'' in
  \emph{IEEE Conference on Computer Communications (INFOCOM)}, 2023.

\bibitem{9796516}
T.~Huang \emph{et~al.}, ``Learning tailored adaptive bitrate algorithms to
  heterogeneous network conditions: A domain-specific priors and
  meta-reinforcement learning approach,'' \emph{{IEEE} J. Sel. Areas Commun.},
  2022.

\bibitem{10.1145/3503161.3548331}
N.~Kan, Y.~Jiang, C.~Li, W.~Dai, J.~Zou, and H.~Xiong, ``Improving
  generalization for neural adaptive video streaming via meta reinforcement
  learning,'' in \emph{MM - Proc. ACM Int. Conf. Multimed.}, 2022, p.
  3006–3016.

\bibitem{10077780}
W.~Li, X.~Li, Y.~Xu, Y.~Yang, and S.~Lu, ``Metaabr: A meta-learning approach on
  adaptative bitrate selection for video streaming,'' \emph{{IEEE} Trans.
  Mobile Comput.}, pp. 1--17, 2023.

\bibitem{ross2010efficient}
S.~Ross and D.~Bagnell, ``Efficient reductions for imitation learning,'' in
  \emph{Proceedings of the 13th International Conference on Artificial
  Intelligence and Statistics (AISTATS)}, 2010, pp. 661--668.

\bibitem{ross2011reduction}
S.~Ross, G.~Gordon, and D.~Bagnell, ``A reduction of imitation learning and
  structured prediction to no-regret online learning,'' in \emph{Proceedings of
  the 14th international conference on artificial intelligence and statistics
  (AISTATS)}, 2011, pp. 627--635.

\bibitem{Yang_Vereshchaka_Zhou_Chen_Dong_2020}
F.~Yang, A.~Vereshchaka, Y.~Zhou, C.~Chen, and W.~Dong, ``Variational
  adversarial kernel learned imitation learning,'' \emph{Proceedings of the
  AAAI Conference on Artificial Intelligence}, vol.~34, no.~04, pp. 6599--6606,
  Apr. 2020.

\bibitem{garg2021iq}
D.~Garg, S.~Chakraborty, C.~Cundy, J.~Song, and S.~Ermon, ``Iq-learn: Inverse
  soft-q learning for imitation,'' in \emph{Neural Information Processing
  Systems (NeurIPS)}, vol.~34, 2021, pp. 4028--4039.

\bibitem{Liu_Liu_Zhao_Pan_Liu_2020}
Y.~Liu, Q.~Liu, H.~Zhao, Z.~Pan, and C.~Liu, ``Adaptive quantitative trading:
  An imitative deep reinforcement learning approach,'' \emph{Proceedings of the
  AAAI Conference on Artificial Intelligence}, vol.~34, no.~02, pp. 2128--2135,
  Apr. 2020.

\bibitem{singh2021parrot}
A.~Singh, H.~Liu, G.~Zhou, A.~Yu, N.~Rhinehart, and S.~Levine, ``Parrot:
  Data-driven behavioral priors for reinforcement learning,'' in
  \emph{International Conference on Learning Representations (ICLR)}, 2021.

\bibitem{10032264}
R.~Huang, V.~W. Wong, and R.~Schober, ``Rate-splitting for intelligent
  reflecting surface-aided multiuser vr streaming,'' \emph{IEEE Journal on
  Selected Areas in Communications}, vol.~41, no.~5, pp. 1516--1535, 2023.

\bibitem{10296872}
D.~Wu, D.~Zhang, M.~Zhang, R.~Zhang, F.~Wang, and S.~Cui, ``Ilcas: Imitation
  learning-based configuration- adaptive streaming for live video analytics
  with cross-camera collaboration,'' \emph{{IEEE} Trans. Mobile Comput.}, pp.
  1--15, 2023.

\bibitem{10026241}
Z.~Ning, H.~Chen, E.~C.~H. Ngai, X.~Wang, L.~Guo, and J.~Liu, ``Lightweight
  imitation learning for real-time cooperative service migration,''
  \emph{{IEEE} Trans. Mobile Comput.}, pp. 1--18, 2023.

\bibitem{LianchenJia}
L.~Jia, C.~Zhou, T.~Huang, C.~Li, and L.~Sun, ``Rdladder: Resolution-duration
  ladder for vbr-encoded videos via imitation learning,'' in \emph{IEEE
  Conference on Computer Communications (INFOCOM)}, 2023.

\bibitem{10258330}
P.~S. Chib and P.~Singh, ``Recent advancements in end-to-end autonomous driving
  using deep learning: A survey,'' \emph{IEEE Transactions on Intelligent
  Vehicles}, pp. 1--18, 2023.

\bibitem{pmlr-v119-siddique20a}
U.~Siddique, P.~Weng, and M.~Zimmer, ``Learning fair policies in
  multi-objective ({D}eep) reinforcement learning with average and discounted
  rewards,'' in \emph{Proceedings of the 37th International Conference on
  Machine Learning}, ser. Proceedings of Machine Learning Research, vol.
  119.\hskip 1em plus 0.5em minus 0.4em\relax PMLR, 13--18 Jul 2020, pp.
  8905--8915.

\bibitem{9837474}
J.~Shao, Y.~Mao, and J.~Zhang, ``Task-oriented communication for multidevice
  cooperative edge inference,'' \emph{{IEEE} Trans. Wireless Commun.}, vol.~22,
  no.~1, pp. 73--87, 2023.

\bibitem{10.1145/3452296.3472923}
A.~Narayanan \emph{et~al.}, ``A variegated look at 5g in the wild: Performance,
  power, and qoe implications,'' in \emph{Proc. Annu. Conf. ACM Spec. Interest
  Group Data Commun. Appl., Technol., Archit., Protoc. Comput. Commun.
  (SIGCOMM)}, 2021.

\bibitem{riiser2013commute}
H.~Riiser \emph{et~al.}, ``Commute path bandwidth traces from 3g networks:
  analysis and applications,'' in \emph{ACM Multimedia Systems Conference
  (MMSys)}, 2013.

\bibitem{fcc}
F.~C. Commission, ``Federal communications commission. 2016. raw data -
  measuring broadband america. (2016),''
  \url{https://www.fcc.gov/reports-research/reports/}.

\bibitem{narayanan2020lumos5g}
A.~Narayanan \emph{et~al.}, ``Lumos5g: Mapping and predicting commercial mmwave
  5g throughput,'' in \emph{ACM Multimedia Systems Conference (MMSys)}, 2020.

\end{thebibliography}
\end{document}